\documentclass[a4paper,11pt]{article}
\usepackage[british]{babel}
\usepackage{color} % colors
\usepackage[table]{xcolor} % colors
\usepackage{transparent}
\usepackage{array}
\usepackage{graphicx} % graphics
\usepackage{stmaryrd}
\usepackage{svg}
\usepackage{tikz} % custom drawing
\usetikzlibrary{shapes,arrows,positioning,automata,backgrounds,calc,er,patterns,decorations.markings, shapes.misc,arrows.meta,positioning,fit,backgrounds} % tikz instances
\usepackage[compat=1.1.0]{tikz-feynman} % feynman diagrams
\usepackage{circuitikz} % draw circuits with latex :-)
\ctikzset{bipoles/length=.8cm}
\usepackage[font={normalsize,normalfont}]{caption}
\usepackage{subcaption} % contains the 'subfigure' environment
\usepackage{physics} % pyhsics symbols
\usepackage{units} % proper units in formulae ( usage: \unit[<value>]{<dimension>} or \unitfrac[<value>]{<numerator>}{<denominator>} )
\usepackage{mathtools} % more symbols
\usepackage{amssymb}
\usepackage{wrapfig} % make inline figures wrapped by text
\usepackage{tabularx}
\usepackage{booktabs}
\usepackage{makecell}
\usepackage{multirow}
\usepackage{multicol}
\usepackage{marginfix}
\usepackage[
    textwidth=2.5cm,
    textsize=tiny
]{todonotes}
\usepackage[
    top=2.5cm,
    bottom=2.5cm,
    left=3cm,
    right=3cm
]{geometry}
\usepackage{adjustbox}
\usepackage{authblk}
\usepackage{orcidlink}
\usepackage{placeins}
\usepackage[backend=biber,
            style=chem-acs,
            articletitle=true,
            maxbibnames=1,
            url=false,
            doi=true,
            isbn=false,
            ]{biblatex}

\AtEveryBibitem{%
  \clearlist{language}%
}
\usepackage{biblatex-shortfields}
\usepackage{lineno}
\usepackage[utf8]{inputenc}
\usepackage[T1]{fontenc}
\newcommand{\ThT}{%
  Tang.\textsc{g}wan\,---\,\allowbreak
  \d{h}a\v{c}x\textsuperscript{w}iqak\,---\,\allowbreak
  Tsi\b{g}is%
}

\usepackage{hyperref}
\hypersetup{colorlinks=true, linkcolor=black, citecolor=black, urlcolor=blue}
\usepackage{cleveref}

\Crefname{tabular}{tab.}{tabs.}
\Crefname{table}{tab.}{tabs.}

\providecommand{\keywords}[1]
{
  \small	
  \textbf{\textit{Keywords -- }} #1
}
\makeatletter
\patchcmd{\@maketitle}
  {\@author}
  {\small\@author}
  {}{}
\makeatother

\newif\ifisoreview
\isoreviewtrue   % true: isolated review version; false: full text
\isoreviewfalse

\makeatletter
\newif\ifiso@inside \newif\ifiso@skipped \newif\ifiso@pending
\newif\ifiso@fresh  \newif\ifiso@drop    \newif\ifiso@mute
\newif\ifiso@remark
\newwrite\iso@out
\newwrite\iso@aux
\newcommand\iso@w[1]{\immediate\write\iso@out{\unexpanded{#1}}}

\ifisoreview
\fi

\newcommand\iso@unskip{%
  \ifnum\lastnodetype=11 \unskip\expandafter\iso@unskip\fi}
\newcommand\gap{\par\medskip{\centering\dots\par}\medskip}
\newcommand\isodots{\leavevmode\unskip\space\dots\space\ignorespaces}
\newcommand\isolead{\leavevmode\dots\space\ignorespaces}

\newcommand\iso@levels{section,subsection,subsubsection}
\newcommand\iso@counters{figure,table,equation,footnote}

\NewDocumentCommand\isolate{+m}{%
  \ifisoreview
    \iso@flushheads
    \ifiso@skipped\ifiso@fresh\else\iso@w{\gap}\fi\fi
    \global\iso@insidetrue
    \ifvmode
      \leavevmode                                  % starts a paragraph
    \else
      \iso@unskip
        \ifdim\lastkern=1sp\else                     % text omitted before it
            \ifiso@pending\iso@w{\isodots}\else\iso@w{\isolead}\fi
        \fi
    \fi
    \iso@synccounters
    \iso@w{#1}%
    #1\kern1sp                                     % end marker
    \global\iso@insidefalse
    \iso@skippedfalse\iso@freshfalse\global\iso@pendingtrue
  \else
    #1%
  \fi}

\AddToHook{para/begin}{%
  \ifisoreview\ifiso@mute\else\ifiso@inside\else\iso@skippedtrue\fi\fi\fi}
\AddToHook{para/end}{%
  \ifisoreview\ifiso@mute\else\ifiso@inside\else\ifiso@pending
    \iso@unskip
    \ifdim\lastkern=1sp \iso@w{\par}\else\iso@w{\isodots\par}\fi
    \global\iso@pendingfalse
  \fi\fi\fi\fi}

\newcommand\iso@pre{%
  \ifisoreview\ifiso@mute\else\ifiso@pending\ifhmode
    \iso@unskip
    \ifdim\lastkern=1sp \global\iso@remarktrue\fi
  \fi\fi\fi\fi}
\newcommand\iso@post{%
  \ifiso@remark\global\iso@remarkfalse\ifhmode\kern1sp\fi\fi}
\newcommand\iso@hideenv[1]{%
  \AddToHook{env/#1/before}{\iso@pre}%
  \AddToHook{env/#1/begin}{\iso@mutetrue}%
  \AddToHook{env/#1/after}{\iso@post}}
\iso@hideenv{figure}\iso@hideenv{figure*}
\iso@hideenv{table}\iso@hideenv{table*}

\newcommand\iso@synccounters{%
  \@for\iso@c:=\iso@counters\do{%
    \@ifundefined{c@\iso@c}{}{%
      \immediate\write\iso@out
        {\string\setcounter{\iso@c}{\the\value{\iso@c}}\@percentchar}}}}

\newcommand\iso@heading[3]{%
  \ifisoreview
    \IfBooleanTF{#2}{\def\iso@num{}}{\edef\iso@num{\the\value{#1}}}%
    \iso@dropfalse
    \@for\iso@l:=\iso@levels\do{%
      \edef\iso@tmp{#1}\ifx\iso@l\iso@tmp\iso@droptrue\fi
      \ifiso@drop\global\expandafter\let\csname iso@pend@\iso@l\endcsname\relax\fi}%
    \global\expandafter\edef\csname iso@pend@#1\endcsname
      {\noexpand\isoheading{#1}{\iso@num}{\unexpanded{#3}}}%
    \iso@skippedfalse\iso@freshtrue
  \fi}
\newcommand\iso@flushheads{%
  \@for\iso@l:=\iso@levels\do{%
    \expandafter\let\expandafter\iso@tmp\csname iso@pend@\iso@l\endcsname
    \ifx\iso@tmp\relax\else
      \immediate\write\iso@out{\unexpanded\expandafter{\iso@tmp}}%
      \global\expandafter\let\csname iso@pend@\iso@l\endcsname\relax
    \fi}}

\newcommand\isoheading[3]{%
  \ifx\relax#2\relax
    \csname #1\endcsname*{#3}%
  \else
    \setcounter{#1}{\numexpr#2-1\relax}\csname #1\endcsname{#3}%
  \fi}

\newcommand\iso@wrap[1]{%
  \expandafter\NewCommandCopy\csname iso@orig@#1\expandafter\endcsname\csname #1\endcsname
  \expandafter\RenewDocumentCommand\csname #1\endcsname{s o m}{%
    \IfBooleanTF{##1}{\csname iso@orig@#1\endcsname*{##3}}%
      {\IfValueTF{##2}{\csname iso@orig@#1\endcsname[##2]{##3}}%
                      {\csname iso@orig@#1\endcsname{##3}}}%
    \iso@heading{#1}{##1}{##3}}}

\def\bibstyle#1{\gdef\iso@bibstyle{#1}}

\newcommand\iso@setupfootnote{%
  \NewCommandCopy\iso@orig@footnote\footnote
  \RenewDocumentCommand\footnote{o +m}{%
    \iso@pre
    \begingroup\iso@mutetrue
      \IfValueTF{##1}{\iso@orig@footnote[##1]{##2}}{\iso@orig@footnote{##2}}%
    \endgroup
    \iso@post}}
\newcommand\iso@immediatewrites{%
  \let\iso@orig@pwrite\protected@write
  \long\def\protected@write##1##2##3{%
    \begingroup\let\thepage\relax##2%
      \let\protect\@unexpandable@protect
      \edef\reserved@a{\immediate\write##1{##3}}\reserved@a
    \endgroup}}
\newcommand\iso@setupbibtex{%
  \begingroup
    \makeatletter
    \let\@writefile\@gobbletwo \let\newlabel\@gobbletwo
    \let\citation\@gobble \let\bibdata\@gobble
    \InputIfFileExists{\jobname-iso.aux}{}{}%
  \endgroup
  \def\bibliography##1{\gdef\iso@bibfiles{##1}}%
  \let\iso@orig@bibstyle\bibliographystyle
  \def\bibliographystyle##1{\gdef\iso@bibstyle{##1}\iso@orig@bibstyle{##1}}}

\AtBeginDocument{%
  \iso@wrap{section}\iso@wrap{subsection}\iso@wrap{subsubsection}%
  \iso@setupfootnote
  \ifisoreview
    \immediate\openout\iso@out=\jobname.iso
    \iso@freshtrue
    \iso@immediatewrites
    \@ifpackageloaded{biblatex}{}{\iso@setupbibtex}%
  \fi}

\AddToHook{shipout/before}{\ifisoreview\DiscardShipoutBox\fi}
\AddToHook{enddocument}{%
  \ifisoreview
    \par\clearpage
    \immediate\closeout\iso@out
    \isoreviewfalse
    \let\protected@write\iso@orig@pwrite
    \setcounter{page}{1}%
    \let\label\@gobble \let\ltx@label\@gobble  % no duplicate labels
    \@ifpackageloaded{biblatex}{%
      \begin{refsection}%
        \input{\jobname.iso}%
        \printbibliography
      \end{refsection}%
    }{%
      \immediate\openout\iso@aux=\jobname-iso.aux
      \let\@auxout\iso@aux
      \immediate\write\@auxout{\relax}%
      \@ifundefined{iso@bibstyle}{}{%
        \immediate\write\@auxout{\string\bibstyle{\iso@bibstyle}}}%
      \input{\jobname.iso}%
      \@ifundefined{iso@bibfiles}{}{%
        \immediate\write\@auxout{\string\bibdata{\iso@bibfiles}}%
        \@input@{\jobname-iso.bbl}}%
    }%
  \fi}
\makeatother
\title{P-ONE-1: the first detector line of the Pacific Ocean Neutrino Experiment}

\date{}

\author[a]{O.~Aghaei\,\orcidlink{0000-0003-1415-6164}}
\author[b]{M.~Agostini\,\orcidlink{0000-0003-1151-5301}}
\author[b]{A.~Alexander Wight\,\orcidlink{0000-0001-9214-6178}}
\author[c]{K.~Ba\c{s}}
\author[a]{N.~Baily}
\author[d,e]{P.~S.~Barbeau}
\author[a]{A.~J.~Baron}
\author[c]{S.~Bash}
\author[c]{C.~Bellenghi}
\author[c]{M.~Boehmer}
% ...

\affil[a]{Ocean Networks Canada, University of Victoria, Victoria, BC, Canada}
\affil[b]{Department of Physics and Astronomy, University College London, Gower Street, London, WC1E 6BT, UK}
\affil[c]{Physik-Department, Technische Universit\"{a}t M\"{u}nchen, D-85748 Garching, Germany}
\affil[d]{Department of Physics, Duke University, Durham, NC, 27708, USA}
\affil[e]{Triangle Universities Nuclear Laboratory, Durham, NC, 27708, USA}
\affil[f]{Department of Physics, Engineering Physics \& Astronomy, Queen's University, Kingston, Ontario K7L 3N6, Canada}
\affil[g]{Department of Physics, Simon Fraser University, 8888 University Drive, Burnaby, BC, Canada, V5A 1S6}
\affil[h]{Department of Physics and Astronomy, Michigan State University, East Lansing, MI, 48824, USA}
\affil[i]{Erlangen Centre for Astroparticle Physics, Friedrich-Alexander-Universit{\"a}t Erlangen-N\"{u}rnberg, D-91058 Erlangen, Germany}
\affil[j]{Department of Physics, Elmhurst University, 190 S. Prospect Ave, Elmhurst, IL, 60126, USA}
\affil[k]{Department of Physics, Drexel University, 3141 Chestnut Street, Philadelphia, PA, 19104, USA}
\affil[l]{Institute of Nuclear Physics, Polish Academy of Sciences, Krakow, Poland}
\affil[m]{Department of Physics, University of Alberta, Edmonton, AB, Canada, T6G 2E1}
\affil[n]{Department of Physics and Astronomy, University of Victoria, 3800 Finnerty Road, Victoria, BC, Canada, V8P 5C2}
\affil[o]{Department of Astronomy and Astrophysics, University of Chicago, 5640 South Ellis Avenue, Chicago, IL, 60637, USA}
\affil[p]{Kavli Institute for Cosmological Physics, University of Chicago, 5640 South Ellis Avenue, Chicago, IL, 60637, USA}
\affil[q]{School of Physics and Center for Relativistic Astrophysics, Georgia Institute of Technology, Atlanta, GA, 30332, USA}

\begin{document}
\maketitle

\newpage
% ABSTRACT
\begin{abstract}
Neutrino telescopes instrument optically transparent media with three-dimensional arrays of photosensors to detect Cherenkov radiation induced by neutrino interactions within or near the instrumented volume. The small cross-sections of the weak interaction and the low neutrino fluxes above TeV energies require vast detector volumes, which only natural media on Earth can realistically provide. Spurred by the detection of the first neutrino sources by the IceCube Neutrino Observatory, a new generation of neutrino telescopes is emerging to increase the global instrumented volume and detection sensitivity. The Pacific Ocean Neutrino Experiment (P-ONE) aims to install such a telescope in the Northeast Pacific Ocean, building on established neutrino telescope design targets in collaboration with Ocean Networks Canada (ONC). This article presents the design, integration, and testing of the first P-ONE line, P-ONE-1, in preparation for its deployment and integration into ONC's deep-sea observatory NEPTUNE. Alongside the detector design and integration methods, we detail in-air and in-water system-level validation performed at several test sites in Canada, culminating in the successful construction of P-ONE-1 and its preparation for deployment.\\[1em]
\noindent \keywords{Neutrino telescope, Deep-sea detector, Detector integration, Data acquisition, Calibration systems}
\end{abstract} 

% TOC
% \newpage
% \setcounter{tocdepth}{2}
\tableofcontents
\noindent\rule{\textwidth}{0.2pt}

%%%%%%%%%%%%%%%%%%%%%%%%%%%%%%%%%%%%%%%%%%%%%%%%
% Introduction
%%%%%%%%%%%%%%%%%%%%%%%%%%%%%%%%%%%%%%%%%%%%%%%%

\section{Introduction}
% - Open with the scientific motivation: neutrino astronomy, astrophysical sources,
%   and the need for a cubic-kilometre scale detector in the Pacific Ocean.
% - Briefly position P-ONE among existing and next-generation neutrino telescopes
%   (IceCube, KM3NeT, Baikal-GVD, TRIDENT).
% - Introduce the staged deployment program: STRAW pathfinder $\to$ P-ONE-1
%   demonstrator $\to$ full array.
% - State the scope and contribution of this paper: the commissioning of the
%   P-ONE-1 demonstrator line, from component production through qualification
%   to deployment readiness.

Neutrino astronomy aims to detect and characterize sources of high-energy neutrinos in the universe. The elusiveness of neutrinos, being solely weakly interacting, and the low expected fluxes above TeV energies make this a challenging task; however, the detection of the first point sources and neutrinos from our Galaxy by IceCube~\cites{TXS.1.2018, Abbasi.2022, GcPlane.2023} and the ultra-high-energy neutrino by KM3NeT~\cite{aielloObservationUltrahighenergyCosmic2025} have demonstrated that it is possible and that a new era of astronomy is accessible with increased detection statistics. 

The main detection signature is Cherenkov light emitted by the relativistic charged secondary particles produced in neutrino interactions. Above the TeV scale, the resulting event morphologies range from cascades of a few tens of meters to muon tracks of several kilometers. Resolving these topologies, while collecting sufficient statistics from the faint astrophysical flux, requires instrumented volumes on cubic-kilometer scales. The challenge is to instrument such a volume densely enough to reject atmospheric and environmental backgrounds, yet sparsely enough to reach the necessary target mass. Moving beyond the discovery of an astrophysical flux towards detailed characterization of source populations and production mechanisms therefore calls for a new generation of detectors that build on the experience of IceCube and KM3NeT while improving particle identification, timing, background suppression, and angular resolution.

The Pacific Ocean Neutrino Experiment (P-ONE)~\cite{NatAst.P-ONE.2020} is a new initiative to achieve greater precision and deep-sea reliability through novel technologies. P-ONE aims to detect astrophysical neutrinos at or above the TeV scale using Pacific Ocean water as the detection medium. With low optical scattering in seawater, the objective is to boost the angular resolution of detected neutrino events by a factor of four compared to IceCube~\cite{abbasiMuontrackReconstructionExploiting2021,twagirayezuPerformancePacificOcean2023}, thus increasing the sensitivity to neutrino point sources~\cite[e.g.~][]{parkCanadianHighEnergyNeutrino2025}. It is also located in the Northern Hemisphere, placing the Galactic Plane and Galactic Center within its sensitive region by detecting up-going neutrinos through the Earth. This reduces atmospheric background, which enhances sensitivity. With this, P-ONE aims to complement the high-energy neutrino sky coverage together with other experimental efforts in the world, among them KM3NeT in the Mediterranean Sea~\cite{LoI.KM3NeT.2016}, GVD in Lake Baikal~\cite{GVD.2023}, built and proposed upgrades of IceCube~\cite{ishiharaIceCubeUpgradeDesign2019,aartsenIceCubeGen2WindowExtreme2020}, and proposed initiatives in the South China Sea~\cite{TRIDENT.Nature.2023,chenHUNTUltralargescaleNeutrino2026,zhangProposedDeepSea2025}.

P-ONE is under construction, starting with the first line, P-ONE-1. The production, testing, and qualification of this first line are the main focus of this article. 

%%%%%%%%%%%%%%%%%%%%%%%%%%%%%%%%%%%%%%%%%%%%%%%%
% P-ONE detector design, overview
%%%%%%%%%%%%%%%%%%%%%%%%%%%%%%%%%%%%%%%%%%%%%%%%

\section{The P-ONE detector}
\label{sec:detector}
% - Science goals: neutrino flux sensitivity, source classes targeted, role of
%   Pacific Ocean site (depth, optical properties, ONC infrastructure).
% - Comparison to existing and next-generation telescopes: IceCube, KM3NeT,
%   Baikal-GVD, TRIDENT --- differentiating factors of P-ONE architecture.
% - STRAW pathfinder: key results and how they informed P-ONE-1 design.
% - Staged deployment roadmap: P-ONE-1 as the demonstrator validating the full
%   technology chain before array-scale production.
% - Sensitivity simulations or expected performance benchmarks if available.

In collaboration with Ocean Networks Canada (ONC)\footnote{Ocean Networks Canada, Canada. \url{https://www.oceannetworks.ca}}, P-ONE leverages existing deep-sea infrastructure and operational expertise. This effort builds on two successful pathfinder missions, STRAW~\cite{Boehmer.2019} and STRAW-b~\cite{Holzapfel.2023b} -- prototype detectors of \unit[150]{m} and \unit[450]{m} lengths -- deployed in 2018 and 2020, respectively. Both targeted the optical characterization of deep-sea water at the future P-ONE site, and their data helped establish the Cascadia Basin abyssal plain as a suitable site for a future neutrino telescope~\cite{Boehmer.2019}. 

P-ONE will be deployed in the Cascadia Basin and integrated into ONC's 800-km-long cabled observatory called \textit{North-east Pacific Time-series Undersea Networked Experiments} (NEPTUNE) in the Pacific Ocean~\cite{NEPTUNE.Web}. This infrastructure provides both power delivery and high-bandwidth network connectivity to the detector elements. The current detector design targets up to 100 detection lines anchored to the seafloor. Each of them comprises twenty optical modules uniformly distributed along a one-kilometer cable kept upright with a subsea float. Multiple detection lines form the full detector array and are interconnected via a network of seafloor junction boxes, which ultimately interface with NEPTUNE for power and data transfer. 

Building, installing, and operating kilometer-tall mooring lines at a water depth of \unit[2.6]{km} is technically challenging and expected to require design iterations, motivating a phased approach. Initially, P-ONE-1 will connect directly to the NEPTUNE node via a wet-mateable cable. Its string junction box (sJB), located at the base of the line, provides the power, network, and timing infrastructure needed to interface with NEPTUNE. Next, the P-ONE Demonstrator (P-ONE-D) will install the first cluster junction box (cJB), which aggregates P-ONE-1 and additional detector lines into a single interface to the NEPTUNE node. Ultimately, a detector junction box (dJB) will interconnect multiple cJBs to the NEPTUNE node, supporting the natural scalability of the array while housing the hardware required for full detector operation. These tiers are visualized in \cref{fig:architecture}.

% %%%%%%%%%%%%%%%%%%%%%%%%%%%%%%%%%%%%%%%%%%%%%%%%
% Detector architecture
% %%%%%%%%%%%%%%%%%%%%%%%%%%%%%%%%%%%%%%%%%%%%%%%%

\subsection{Detector architecture}
\label{subsec:pone:architecture}

\begin{figure}[h!]
    \centering
    \includegraphics[trim=0 0 0.5cm 0,clip,width=.99\textwidth]{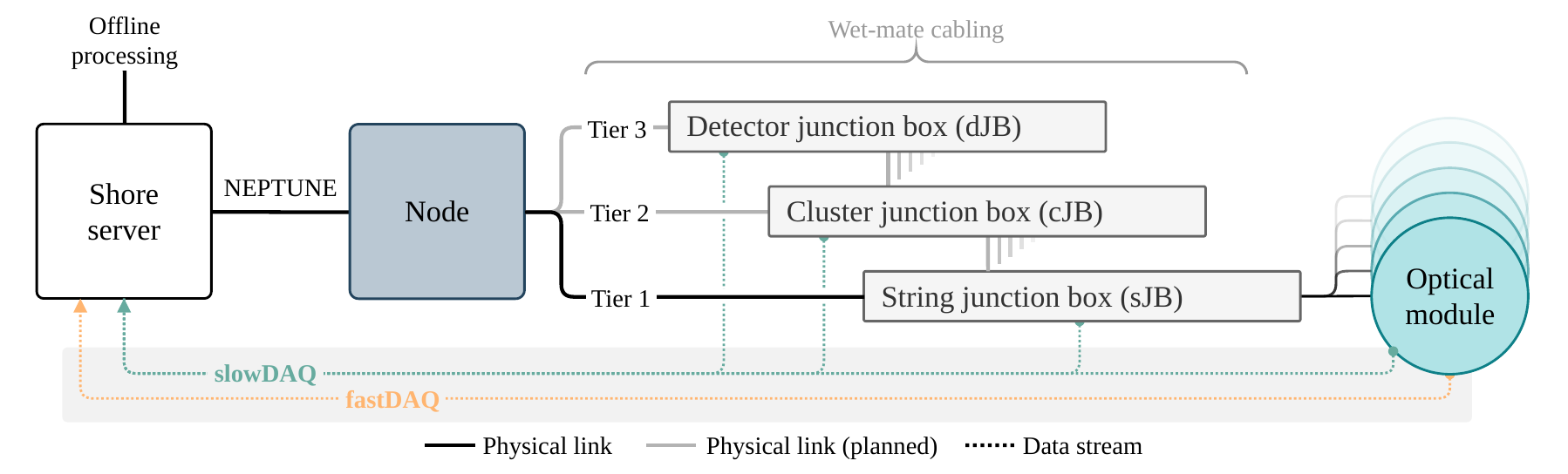}
    \caption{Planned tiers of the P-ONE detector architecture within the ONC network. P-ONE-1 forms the first, self-consistent tier compatible with future superseding infrastructure.}
    \label{fig:architecture}
\end{figure}

The data acquisition (DAQ) system operates as a network-based push-pull architecture controlled from a central host (\cref{fig:architecture}). From the node, data are distributed over a passive optical network via fiber-optic TCP/IP and routed across junction box tiers via commercial-grade optical switches and splitters. Optical modules and junction box subsystems connect as network clients, enabling direct configuration and data transfer. Timing is distributed independently over fiber using custom hardware to achieve picosecond precision~\cites{Bohmer.ICRC23,daq_inprogress}, and each new detector tier supplies a root timing source that supersedes the upstream clock for array-wide synchronization. The total available bandwidth allocated for P-ONE is \unit[1]{Gigabit per second (Gbps)}, and upgrades are planned to extend this to \unit[10]{Gbps}.
\par
Experiment control is implemented through slow-control software organized into three tiers of code packages. At the lowest level, the \textit{perseus} code base collects device-specific drivers for all module and junction box hardware, providing direct network-based hardware access for control and read-back via the \textit{tuberd} network-interfacing codebase~\cite{tuberd}. At the highest level, the \textit{Maximum Integrated Data Acquisition System} (MIDAS)~\cite{midaswiki} on the central shore server coordinates detector and run configuration, data taking and storage, and exposes a high-level web interface for detector operation. Bridging these two layers, the \textit{andromeda} code package abstracts the perseus hardware interface into MIDAS-accessible functions covering specific detector functionalities and sequencing, and web frontends.

% %%%%%%%%%%%%%%%%%%%%%%%%%%%%%%%%%%%%%%%%%%%%%%%%
% Design overview
% %%%%%%%%%%%%%%%%%%%%%%%%%%%%%%%%%%%%%%%%%%%%%%%%

\subsection{P-ONE-1: Design overview}
\label{subsec:pone:overview}
% - Line architecture overview: diagram showing the full line with P-OMs,
%   P-CAL modules, acoustic modules, SJB, cable segments, and surface float.
% - P-OM design: pressure housing, PMT count and type, mainboard, MIST,
%   flasher, acoustic transceiver. Key design drivers: optical performance,
%   pressure/depth rating, noise budget, calibration capability.
% - Calibration and auxiliary subsystems: P-CAL, acoustic positioning,
%   SJB functionality, ONC infrastructure compatibility.
% - Component provenance as a consequence of design: each institution
%   contributed specialised expertise that shaped its component.
%     - Interposer (UoA), PMTs (Hamamatsu), microbases and 3D inserts (TUM),
%       glass hemispheres (Nautilus).
%     - Flashers and acoustic modules (SFU), mainboards (MSU), SJB (ONC),
%       acoustic transceivers (Sonardyne), cables (MacArtney).
% - Incoming acceptance checks at receiving sites; convergence at TRIUMF
%   as the integration hub.

\begin{figure}[h!]
    \centering
    \includegraphics[width=0.9\textwidth]{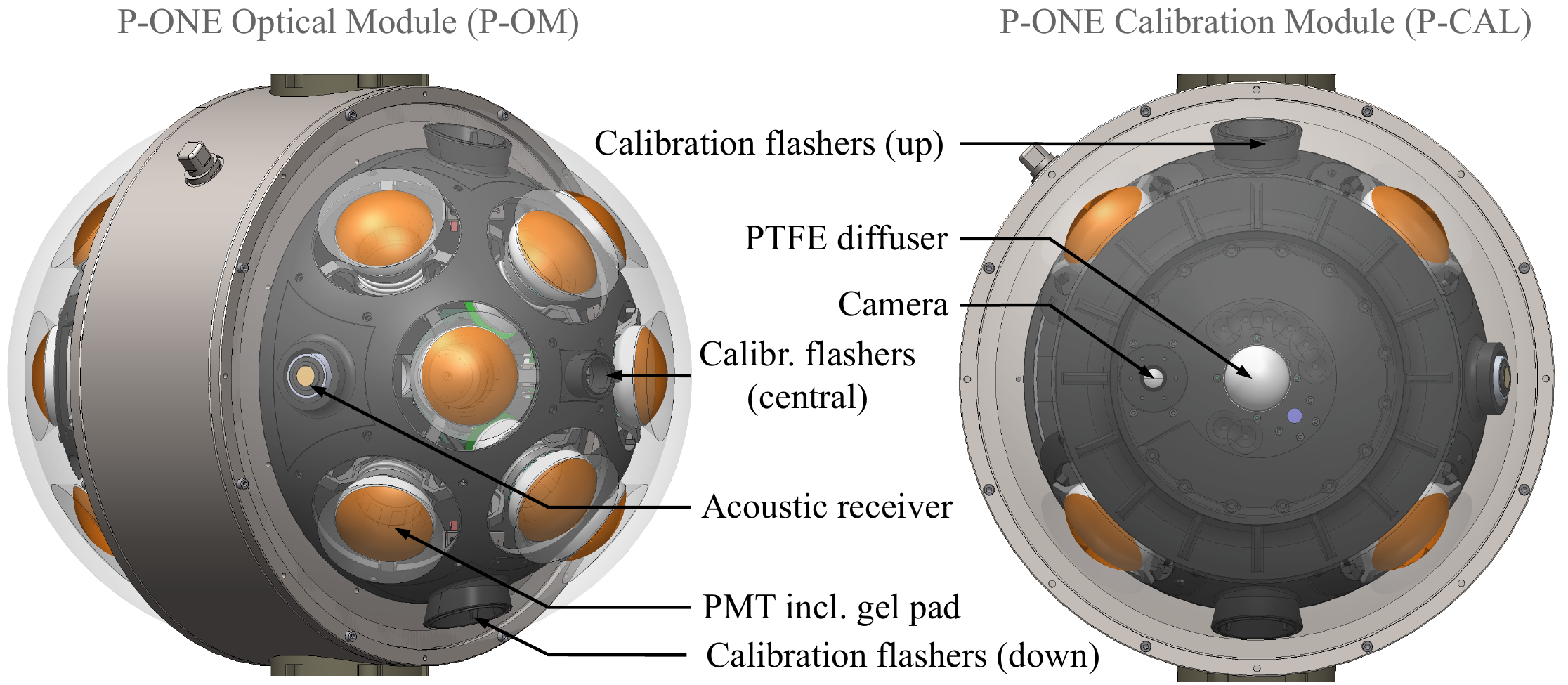}
    \caption{Isometric view of an optical module (left) and frontal view of a calibration module (right). Each optical module hemisphere carries eight PMTs, one acoustic receiver, and a set of directional flashers, with the flashers pointing up, down, and perpendicular to the line. Calibration module hemispheres replace four PMTs with a diffuse flasher system and a camera system.}
    \label{fig:design}
\end{figure}

The P-ONE-1 design comprises twenty detection modules distributed along a one-kilometer detector line: 18 optical modules (P-OMs)~\cite{Spannfellner.2026}, each instrumented with 16 PMTs, one calibration module (P-CAL)~\cite{OptCal.2026} with 8 PMTs, and one hybrid module pairing a P-OM hemisphere with a P-CAL hemisphere; both hemisphere designs are shown in \cref{fig:design}. All modules are encapsulated in a 17" glass pressure housing connected to a titanium flange produced by Nautilus Marine Service GmbH\footnote{Nautilus Marine Service GmbH, Germany. \url{https://www.nautilus-gmbh.com}}, and selected modules are polished or coated with ClearSignal\texttrademark{} anti-biofouling coating. The reduced PMT count in the P-CAL hemispheres frees up space for diffuse flasher and camera systems, with the P-CAL and hybrid module positioned at approximately 300 and \unit[650]{m} from the seafloor, respectively.
\par
\begin{figure}[h!]
    %\hspace*{0.1\textwidth}%
    \centering
    \includegraphics[width=0.85\textwidth]{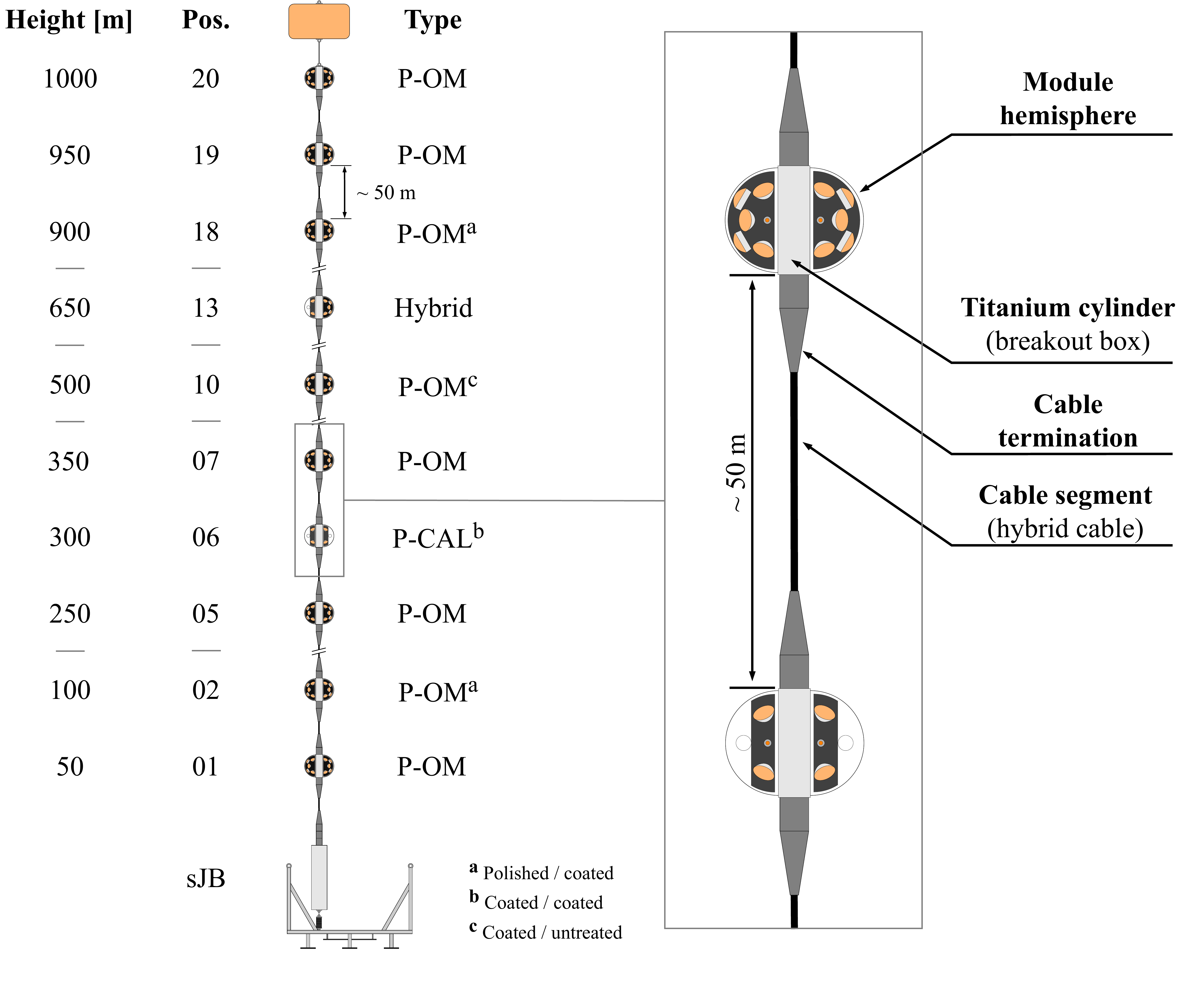}
    \vspace*{-2em}
    \caption{Conceptual drawing of the P-ONE-1 line showing 10 of 20 modules (left) and a detailed module segment view (right). Calibration module (P-CAL) hemispheres are located at positions 06 and 13, corresponding to $300\,$m and $650\,$m above the seafloor, respectively. Superscripts indicate which hemisphere treatments are present at each position. Figure adapted from~\cite{Spannfellner.2026}.}
    \label{fig:line}
\end{figure}

A custom cable assembly combines copper conductors, optical fibers, and load-bearing elements in one line with a diameter of about \unit[32]{mm}. A connector-less design eliminates pressure housing penetrators, one of the most failure-prone points in deep-sea systems. The resulting design builds on three independent seals to protect against water ingress into the optical modules and cable. At predefined module positions, a subset of fibers and conductors branch out and terminate in a titanium cylinder that mates with the module's titanium flange, integrating the module pressure housing into the line both mechanically and functionally. The cable assembly was manufactured by MacArtney A/S\footnote{MacArtney A/S, Denmark. \url{https://www.macartney.com}}. The line terminates at the sJB, and a subsea float at the top holds it upright in the water column. The resulting line layout is shown in \cref{fig:line}.

\begin{figure}[h!]
    \centering
    \begin{subfigure}[t]{0.45\textwidth}
        \centering
        \includegraphics[width=0.75\textwidth]{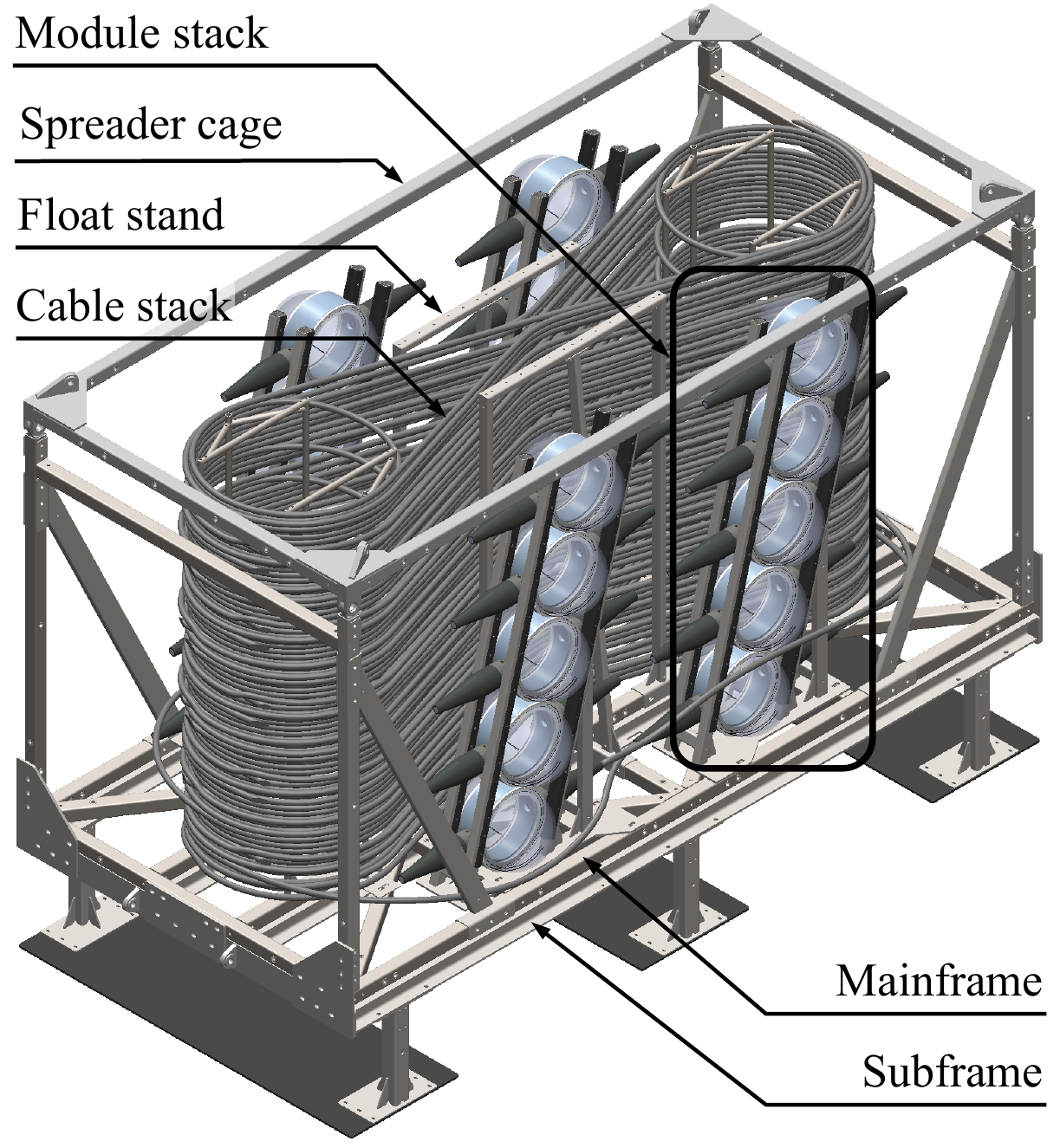}
        \caption{Deployment frame CAD drawing.}
        \label{fig:deployment-frame-a}
    \end{subfigure}
    \hfill
    \begin{subfigure}[t]{0.54\textwidth}
        \centering
        \includegraphics[width=0.75\textwidth]{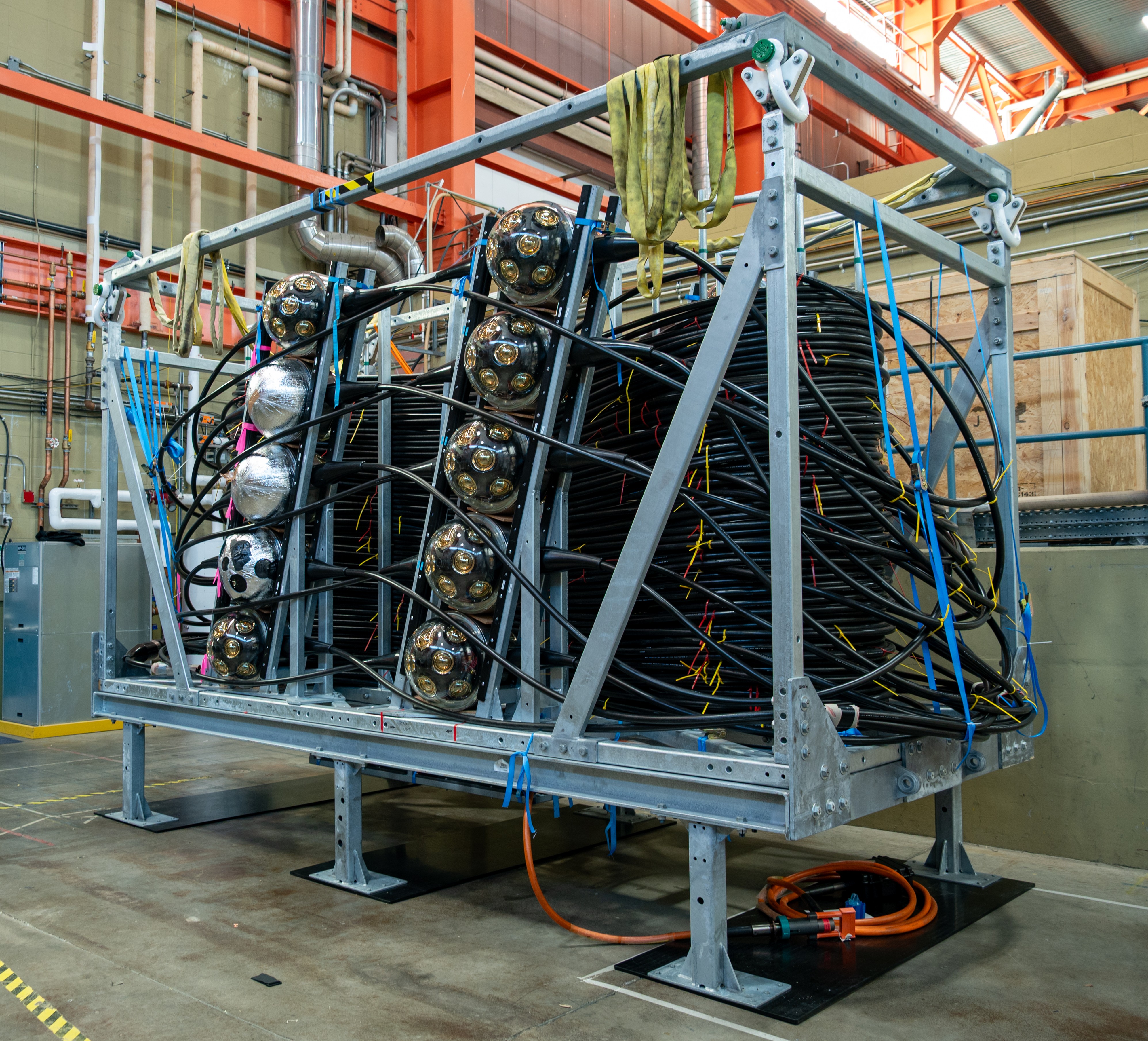}
        \caption{Assembled P-ONE-1 line at TRIUMF.}
        \label{fig:deployment-frame-b}
    \end{subfigure}
    \caption{(\subref{fig:deployment-frame-a}) Rendering of the P-ONE-1 deployment frame with the cable and all optical modules, divided into four separate stacks. The sJB rests in the subframe, and the float is attached at the top before deployment. Once on the seafloor, the float is released, unfurling the line; taken from~\cite{Spannfellner.2026}. (\subref{fig:deployment-frame-b}) Fully assembled P-ONE-1 line at TRIUMF with two visible module stacks.}
    \label{fig:deployment-frame}
\end{figure}

The cable is spooled onto a $4.8 \times 2.3 \times 3.6\,$m deployment frame, comprised of the subframe hosting the sJB and the mainframe hosting the cable and optical modules. The deployment frame serves as a multi-purpose platform for the bottom-up deployment strategy, anchoring the line to the seafloor, and transport; see section \cref{subsec:subsea_install} for details on subsea installation and deployment. Subsea longevity is ensured by a hot-dip galvanized frame and additional sacrificial anodes, which provide galvanic cathodic protection against corrosion. The cable system is manufactured and coiled onto the mainframe by MacArtney A/S in Denmark. Individual module subsystems are assembled at P-ONE institutes and shipped to the final integration site at TRIUMF in Vancouver, Canada. There, the module hemispheres and sJB are integrated on the cable system. A rendering and photograph of the integrated line are shown in \cref{fig:deployment-frame}.

The sJB and module termination were each independently pressure-tested in accordance with the ISO 13628-6:2008-01~\cite{ISO13628-6} standard, successfully withstanding multiple pressure cycles up to the required \unit[325]{bar} and deployment pressure profiles.
\par
%

% %%%%%%%%%%%%%%%%%%%%%%%%%%%%%%%%%%%%%%%%%%%%%%%%
% sJB
% %%%%%%%%%%%%%%%%%%%%%%%%%%%%%%%%%%%%%%%%%%%%%%%%

\subsection{String junction box}
\label{subsec:pone:sjb}

The sJB hosts the necessary infrastructure for connection to NEPTUNE and operation of the line. For the former, a commercial-grade optical uplink switch and multiple power converters connect to the ONC network and the backbone power supply. A series of subordinate power controllers provide controllable power supplies for each optical module and all the DAQ systems in the sJB.

Central data distribution is handled by \textit{BlackCat}~\cite{Bohmer.ICRC23}, a custom protocol that combines Gigabit Ethernet and picosecond timing in a single modified Ethernet link, with one fiber per optical module. On the backend, BlackCat terminates at the optical uplink switch, which provides the connection to the ONC network. As P-ONE-1 is the first P-ONE detector line at the Cascadia node of NEPTUNE, its sJB also houses a root timing source~\cite{SyncBox.DS} and a passive optical timing system. The latter is the default timing path and reaches each optical module over a dedicated fiber through a passive optical splitter. Each optical module is therefore served by two fibers: one for data and timing over BlackCat, and one for passive timing. A second pair of root fibers at the sJB, part of the ONC backbone connection, provides compatibility with the next infrastructure tier. Both BlackCat and the passive timing system accept an external root clock, allowing seamless integration once higher-level infrastructure supersedes the present configuration.

\isolate{Beyond central data and timing distribution, the sJB hosts six single-board Odroid-N2+ computers~\cite{OdroidN2.Web}, a serial server for communication with serial devices, and custom control electronics for the acoustic positioning system. The latter enables active control and time synchronization of the external acoustic interrogator, which is deployed together with P-ONE-1 as part of the broader acoustic positioning system and connects directly to the sJB; see~\cref{subsec:line-testing:mtc}.} Lastly, a multitude of temperature sensors are distributed throughout the sJB's interior, enabling continuous monitoring of the internal thermal environment.

% %%%%%%%%%%%%%%%%%%%%%%%%%%%%%%%%%%%%%%%%%%%%%%%%
% Module components
% %%%%%%%%%%%%%%%%%%%%%%%%%%%%%%%%%%%%%%%%%%%%%%%%

\subsection{Module components}
\label{subsec:pone:module}

Each module consists of two hemispheres, A-type and B-type, which are individually attached to the line. Both hemisphere types share a common baseline, including PMTs, an electronics fan-out board, and calibration devices, all integrated into an internal mechanical mounting structure. This allows standardized production and testing, as well as interchange of hemispheres before final integration with the line.

\begin{figure}[h!]
    \centering
    \includegraphics[width=0.9\textwidth]{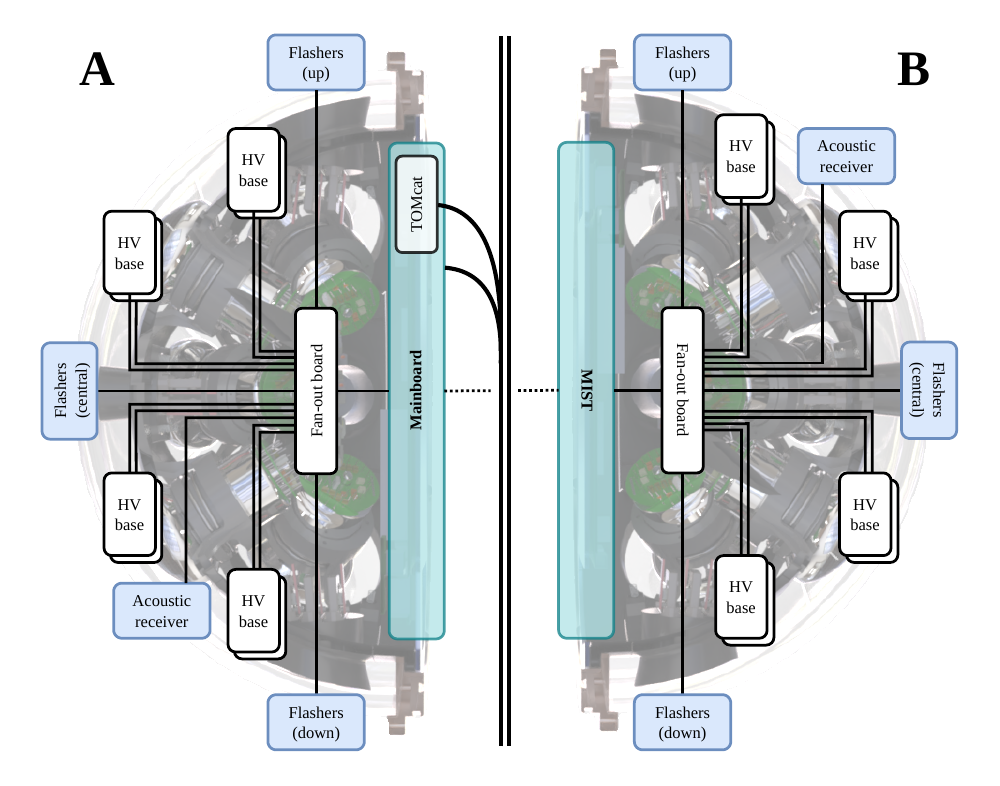}
    \caption{Layout of the module hemisphere systems in P-ONE-1, shown for a regular P-OM. Note that acoustic receivers face into and outside the figure in A- and B-type hemispheres, respectively.}
    \label{fig:module-systems}
\end{figure}

The hemisphere types differ in their central payload: A-type hemispheres are equipped with the module mainboard, providing the primary DAQ infrastructure via an on-board Linux operating system running on the central FPGA, along with all electronic components for waveform digitization, network connectivity, and timing~\cite{daq_inprogress}; B-type hemispheres carry the \textit{Muon In-Situ Tracker} (MIST), made of four scintillator plates instrumented with SiPMs for independent inter-module muon detection~\cite{mist_inprogress}. A summary is shown in \cref{fig:module-systems}.
\par
The mainboard serves as the central interface to the line, receiving power, network, and time synchronization through the module termination's copper wires and optical fibers. A dedicated add-on board, called \textit{TOMcat}, serves as the endpoint for decoding the hybrid data and timing protocol used by the BlackCat system. The fan-out board sits at the center of each hemisphere, connecting all PMTs and calibration subsystems and forming a single-hemisphere connection to the mainboard. This provides two distinct advantages: hemispheres can be pre-assembled and remain fully interchangeable before the integration of a payload (i.e., mainboard or MIST system), and the integration procedure of a module to the line reduces to connecting two fan-out board control cables and two PMT analog cable bundles to the mainboard for each hemisphere. The MIST payload in B-type hemispheres connects to the hemisphere's fan-out board for power and communication.

The PMTs form the core of the light detection system. They are integrated into a vacuum-cast mounting frame that includes mechanical interfaces for all PMTs and four modular slots for integrating other subsystems using a generic mechanical interface. Optical coupling of the PMTs to the glass hemisphere is achieved with custom gel pads molded onto each PMT before integration into the hemisphere. A programmable high-voltage (HV) supply is provided to each PMT using a HV base adapted from the IceCube Upgrade~\cite{mDOM.Proc.2021}, and is controlled via cabled UART serial connections through the fan-out board. The analog outputs of all PMTs are fed directly to a dedicated ADC~\cite{ADC.MB.AD9083} on the mainboard via coaxial cables, bundled and carefully routed through the hemisphere to minimize noise pickup. The sampling rate for PMT waveforms is $208.3\,$MHz, and the readout is controlled through the fastDAQ stream. 

Optical calibration is achieved using dedicated light pulsers, and a detailed description of the P-ONE-1 calibration systems can be found in \cite{OptCal.2026}. A total of 296 directional flashers are integrated into P-ONE-1, with standard P-OMs containing eight flashers per hemisphere individually cabled to the fan-out board. They employ light-emitting diodes and laser diodes with central wavelengths of $365-520\,$nm, chosen to match the expected water transmission~\cite{Bailly.2021}. By pointing them upwards and downwards, the intermediate water column between modules can be calibrated. P-CALs replace four PMTs with a diffuse, self-monitoring light pulser system, and a wide-angle camera operated via USB to monitor sedimentation on the glass hemisphere~\cite{Sedimentation.2026} and bioluminescence~\cite{Li.2026}. Both diffuse and directional flashers will also be used for time calibration. 

For positioning, each hemisphere includes a piezo-acoustic receiver with a programmable-gain amplifier; a detailed description and characterization of this system can be found in~\cite{AcousticCal.2025}. These receivers detect acoustic positioning pings from independent acoustic emitters on the seafloor and enable multilateration of the module's position. Mechanically, the acoustic receiver unit is integrated into one of the generic mounting frame slots and attached to the glass hemisphere using two-part epoxy; electrically, it is connected to the fan-out board for control and readout. The differential output signals are fed to the mainboard, where a dedicated audio ADC~\cite{AudioADC.2019} samples the recorded signals at a baseline frequency of 192\,kHz and provides additional digital features to enhance signal quality. The ADC output data are losslessly encoded in FLAC format~\cite{rfc9639} on-board the module and transmitted to the central DAQ system on shore via network connection.

% %%%%%%%%%%%%%%%%%%%%%%%%%%%%%%%%%%%%%%%%%%%%%%%%
% Data flow
% %%%%%%%%%%%%%%%%%%%%%%%%%%%%%%%%%%%%%%%%%%%%%%%%

\subsection{Data flow}
\label{subsec:pone:data-flow}

Data flow in P-ONE differentiates between two distinct streams: slow and fast. Both are configured through the three-tiered software stack (comprising MIDAS, andromeda, and perseus) but operate independently. The hierarchy and interfaces of this stack are conceptually shown in \cref{fig:code-stack}.
\isolate{
    \begin{figure}[h!]
        \centering
        \includegraphics[width=0.99\textwidth]{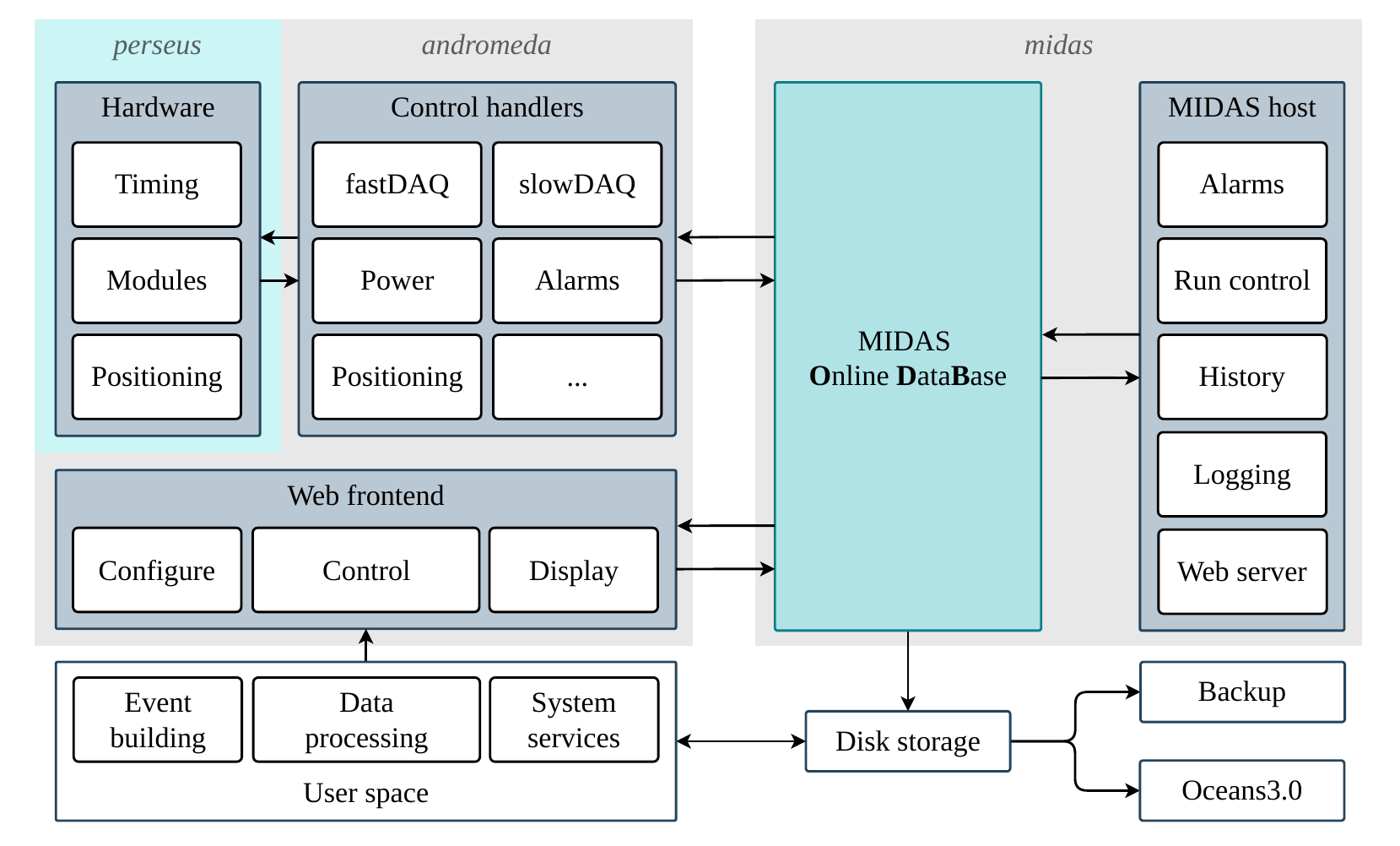}
        \caption{Current workflow and interfaces for detector operation with the MIDAS, andromeda, and perseus stack. The MIDAS online database (ODB) is the central control interface.}
        \label{fig:code-stack}
    \end{figure}\par
}
As part of the slow stream, the andromeda \textit{slowDAQ} handler queries and parses slow data, encompassing all system and sensor read-backs -- including environmental (temperature, pressure, humidity) and orientation (acceleration, magnetic field) sensors, and device configurations. It uses only a small fraction of available bandwidth. The slow stream also carries acoustic waveforms from the piezo receivers in all modules and intermittent camera images from P-CAL modules. Both are acquired via dedicated handlers and together use about \unit[10]{\%} of the available bandwidth.

The fast stream reserves the remaining bandwidth for digitized photomultiplier tube (PMT) waveforms. For this, the \textit{fastDAQ} framework runs dedicated handlers on the optical modules and onshore, forming the basis of the push-pull readout system; see~\cite{daq_inprogress} for a description. On each optical module, the fastDAQ daemon initiates an analog-to-digital converter (ADC) readout of PMT waveforms once a threshold-based local trigger condition is met, optionally requiring a local coincidence condition, such as multiple PMTs recording a hit within a module and a defined time window. Digitized waveforms are buffered on the module for up to 0.8 seconds, while compressed trigger packets -- encoding pulse count and timestamps in 64-bit words -- are continuously forwarded to shore. The shore-side fastDAQ handler evaluates these against the trigger mode and, if a full readout is warranted, reads waveforms from all relevant modules before their buffers expire. This is estimated to allow array readout rates up to \unit[320]{kHz} at \unit[1]{Gbps} bandwidth.

MIDAS centrally configures, schedules, and monitors both the slow and fast data streams, and allows manual and automated interventions as necessary. Based on the detector's operational mode, different schedules and configurations of all data streams are set. A central online database (ODB) maintained by the MIDAS host serves as the primary interface point for detector configuration. Data is stored on disk and periodically archived for offline processing, with a subset sent to ONC's public open-data portal Oceans3.0\footnote{Ocean Networks Canada: Oceans3.0. \url{https://data.oceannetworks.ca}}.

%%%%%%%%%%%%%%%%%%%%%%%%%%%%%%%%%%%%%%%%%%%%%%%%
% TRIUMF production
%%%%%%%%%%%%%%%%%%%%%%%%%%%%%%%%%%%%%%%%%%%%%%%%

\section{Integration at TRIUMF}
\label{sec:triumf}

The production of modules and sJB subsystems was distributed among institutes of the P-ONE collaboration. The final assembly of modules, as well as their integration with the detector line and the sJB integration, was performed at TRIUMF. There, a dedicated cleanroom provided the dust-free environment, space, infrastructure, and tooling required for subsequent integration steps.

Additionally, a server rack enabled the sJB to connect to a network architecture identical to NEPTUNE, emulating the offshore environment and allowing software validation for online operation. This also hosts Dynamic Host Configuration Protocol (DHCP) and Domain Name System (DNS) servers, configuring network settings for all modules and network devices within the sJB. Since both optical modules and the six Odroid-N2+ located in the sJB are configured to boot their operating system (OS) over network, the server also hosts a Trivial File Transfer Protocol (TFTP) server to provide the necessary boot files when powering devices. This makes OS updates significantly more convenient once deployed.

%%%%%%%%%%%%%%%%%%%%%%%%%%%%%%%%%%%%%%%%%%%%%%%%
% Pre-production testing
%%%%%%%%%%%%%%%%%%%%%%%%%%%%%%%%%%%%%%%%%%%%%%%%

\subsection{Pre-production testing}
\label{subsec:pre-prod:test}

Before the full production volume of modules was finalized, a full prototype hemisphere was assembled from production-grade components at TRIUMF. This ensured both optimized integration procedures for all subsystems and the module, and a detailed end-to-end characterization of the module's performance.

%%%%%%%%%%%%%%%%%%%%%%%%%%%%%%%%%%%%%%%%%%%%%%%%
% Noise characterization
%%%%%%%%%%%%%%%%%%%%%%%%%%%%%%%%%%%%%%%%%%%%%%%%

\subsubsection{Prototype noise characterization}
\label{subsec:triumf:noise}
% - First complete P-OM assembly at TRIUMF: integration of sub-components
%   from distributed production sites and lessons learned.
% - Identification of noise sources (EMI, dark noise, microphonics).
% - Mitigation strategies: shielding, grounding, mechanical damping,
%   firmware-level filtering.
% - Validation of final noise performance against requirements.

The critical science requirement for the optical modules is a low noise level on the PMT signal traces, as this noise directly sets the photoelectron detection threshold at the ADC. The P-ONE efficiency requirement for photoelectron detection translates to a maximum standard deviation of about 0.8 ADC counts, essentially matching the intrinsic noise floor of the ADC on the mainboard with no signal cables connected. However, taking PMT data with the first fully integrated prototype module revealed significant noise across multiple frequency bands between \unit[300]{kHz} and \unit[100]{MHz}, resulting in an overall noise level that exceeded defined requirements. Careful investigation of internal and external noise contributions identified the dominant noise source as an internal ground loop between the PMT HV base and the mainboard, which picked up switching noise from internal power supplies through the coaxial cable shield. 

\begin{figure}[h!]
    \centering
    \begin{subfigure}[b]{\textwidth}
        \centering
        \includesvg[width=\textwidth]{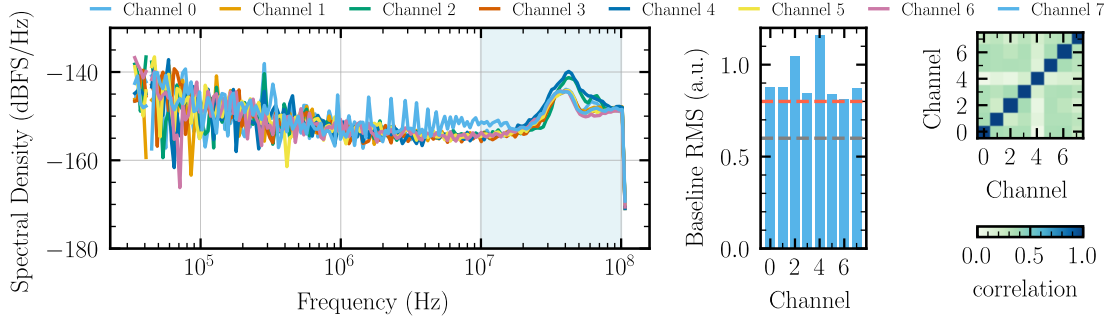}
        \caption{Without noise mitigation.}
        \label{fig:prototype-noise-psd_before}
    \end{subfigure}
    \\[1em]
    \begin{subfigure}[b]{\textwidth}
        \centering
        \includesvg[width=\textwidth]{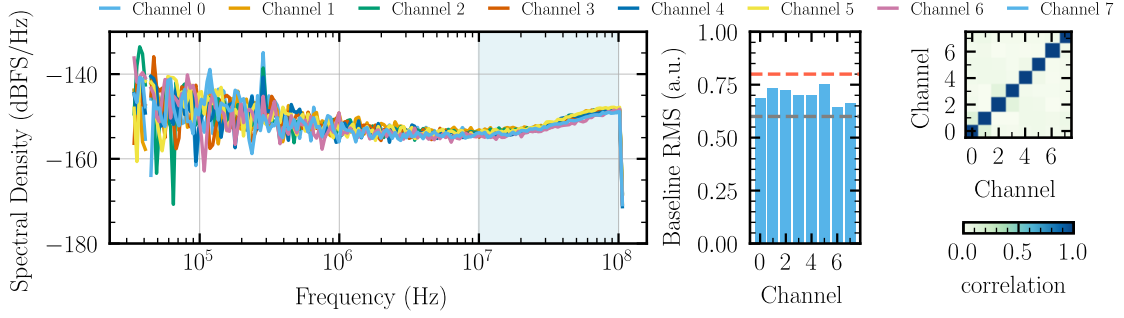}
        \caption{With all noise mitigation strategies implemented.}
        \label{fig:prototype-noise-psd_after}
    \end{subfigure}
    \caption{Baseline properties of the PMT signal lines in a P-ONE-1 hemisphere, showing power spectral density (PSD, left), root-mean-square (RMS, center), and channel-to-channel correlation (right) with and without noise mitigation strategies implemented and PMTs disabled. Also shown are the PMT signal region (blue shading) in the PSD plot, and the noise requirement (red dashed) and ADC noise floor (grey dashed) in the RMS plot.}
    \label{fig:prototype-noise-psd}
\end{figure}

Mitigating this required a multi-step revision process: (a) changing a resistor on the HV base, (b) installing common-mode chokes on the UART communication cables (between the HV bases and the fan-out board), the PMT coaxial signal cables, and the power wires, and (c) realizing a reproducible cable-routing scheme for the PMT signal cables to strictly control impedance. We identified these strategies through an iterative process of evaluating the noise profile from baseline ADC data (i.e., with disabled PMTs but connected signal cables). The pre- and post-mitigation power spectral densities (PSDs) are shown in \cref{fig:prototype-noise-psd_before,fig:prototype-noise-psd_after} and highlight the impact of the ground loop on signal-line noise levels. After implementing all mitigation measures, both hemispheres measure identical noise standard deviations of about \unit[$0.6 - 0.76$]{counts}, indicating a negligible effect of internal noise pickup and meeting all electrical and functional requirements. 
\par
%

%%%%%%%%%%%%%%%%%%%%%%%%%%%%%%%%%%%%%%%%%%%%%%%%
% Vibration and shocks
%%%%%%%%%%%%%%%%%%%%%%%%%%%%%%%%%%%%%%%%%%%%%%%%

\subsubsection{Vibration and shock qualification}
\label{subsec:triumf:vib}

In addition to functional checks, both the sJB and several fully-integrated prototype hemispheres were qualified against vibration and shock to assess system integrity throughout transport and handling.

Three testing standards for classification were considered: MIL-STD-810H~\cite{IEST.MIL-810H}, MIL-STD-167-1A~\cite{MIL-167A}, and ISO 13628-6:2006~\cite{ISO13628-6}. MIL-STD-810H specifies Category 4 Secured Cargo Vibration testing for land transport, consisting of 30-minute random, broadband vibrations in three axes; MIL-STD-167-1A describes an exploratory vibration test for naval transport sweeping frequencies of \unit[5-33]{Hz} with a peak-to-peak amplitude of \unit[0.02]{in} and a sweep rate of \unit[0.067]{Hz/s}. Both MIL standards are specified for land and naval transport, and passing them is strictly required for all P-ONE-1 hardware. We considered the third optional standard ISO 13628-6:2006, which is specified for subsea production control systems in the oil and gas industry. It includes a vibration resonance search, a frequency sweep from \unit[3-150]{Hz} with amplitudes up to \unit[5]{g} (half-sine), and \unit[10]{g} shocks over 11 ms (half-sine). All tests are applied along three axes.

Prototype hemispheres were available for both the P-OM and the P-CAL, allowing them to be tested against the more demanding ISO 13628-6:2006 standard~\cite{ISO13628-6}. The P-CAL prototype, integrating the flasher and acoustic calibration systems, the fan-out board, and a MIST payload, passed with all systems intact and functional. The first P-OM prototype, equipped with a mainboard, developed a mechanical resonance at \unit[47]{Hz} that damaged the mounting system and two large input capacitors. A revised mounting assembly with reinforced bolts and a capacitor brace was introduced to limit resonance-induced stresses. The reworked hemisphere passed a second cycle against the required MIL-STD-167-1A and MIL-STD-810H standards plus an additional \unit[5]{g} shock cycle, with all systems intact and noise performance unchanged. Additionally, a mechanical prototype with the same reinforcements also passed the full ISO test.

\begin{figure}[h!]
    \centering
    \includesvg[width=\textwidth]{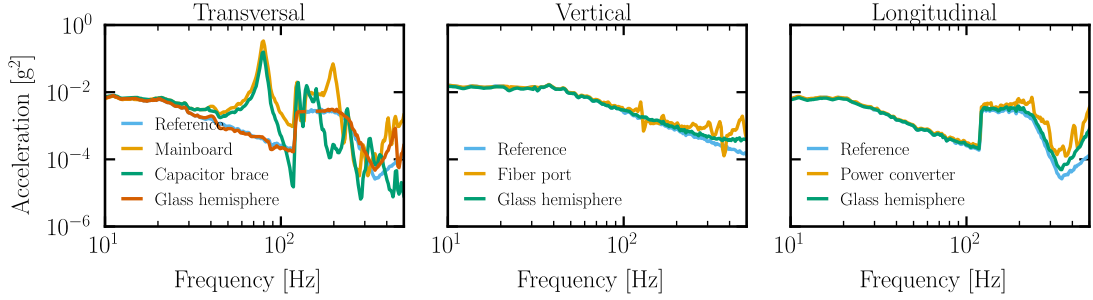} % Do not include .svg extension
    \caption{Results of the MIL-STD-810H Category 4 vibration test for a P-OM prototype hemisphere with reinforced payload mounting infrastructure. Data was collected from several accelerometer probes placed at key structural points: on the excitation table (reference); the center of the mainboard; the apex of the glass hemisphere; and individual probes on the capacitor brace, fiber port, and power converter. The transversal direction (perpendicular to the mainboard plane) shows resonances consistent with the damage mechanism suspected from the first test.}
    \label{fig:vib-random}
\end{figure}

\Cref{fig:vib-random} shows the acceleration measured at several locations within the reinforced P-OM hemisphere during broadband random excitation per MIL-STD-810H. Clear resonances appear in the transversal direction, perpendicular to the mounting plane, consistent with the suspected damage mechanism. The hemisphere nevertheless remained fully functional, suggesting the reinforcements were effective.
\par
The sJB was tested against both required MIL standards to validate system robustness against the anticipated transport and handling processes. Both testing procedures were passed, with all subsystems remaining functional.

%%%%%%%%%%%%%%%%%%%%%%%%%%%%%%%%%%%%%%%%%%%%%%%%
% Module integration
%%%%%%%%%%%%%%%%%%%%%%%%%%%%%%%%%%%%%%%%%%%%%%%%

\subsection{Module integration}
\label{subsec:module-integration}
% - Serial production workflow for all 20 P-OMs: 3D insert mounting,
%   internal cabling, mainboard and MIST integration, hemisphere closure.
% - Per-module QA checklist: visual inspection, continuity, pressure test,
%   HV ramp, baseline dark rate.
% - Traceability: component-level serialisation and test record database.
% - Yield and any remediation actions.

Hemispheres arrived at TRIUMF pre-assembled and functionality-tested with eight PMTs, central flashers, and the fan-out board integrated into the internal mounting frame. Before subsystem integration, a visual inspection of all received sub-components and pre-assembled hemispheres was performed. Afterwards, the up- and down-facing calibration flashers, the acoustic receivers, and identified noise-mitigation strategies were integrated. For each step, we followed a written procedure and performed an independent, two-person quality control. If it passed, the payload was integrated, and the module proceeded to the final acceptance test before being prepared for line integration. 

Because components are produced across various institutes and only fully assembled at TRIUMF, a final acceptance test is essential to verify that all subsystems meet specifications once integrated into a module. For P-ONE-1, we developed a dedicated test system -- the \textit{Module Integration and Network Test} (MINT) -- serving as the final acceptance stage to confirm that all subsystems in a given module operate as expected.

\begin{figure}[h!]
    \centering
    \includegraphics[width=0.9\textwidth]{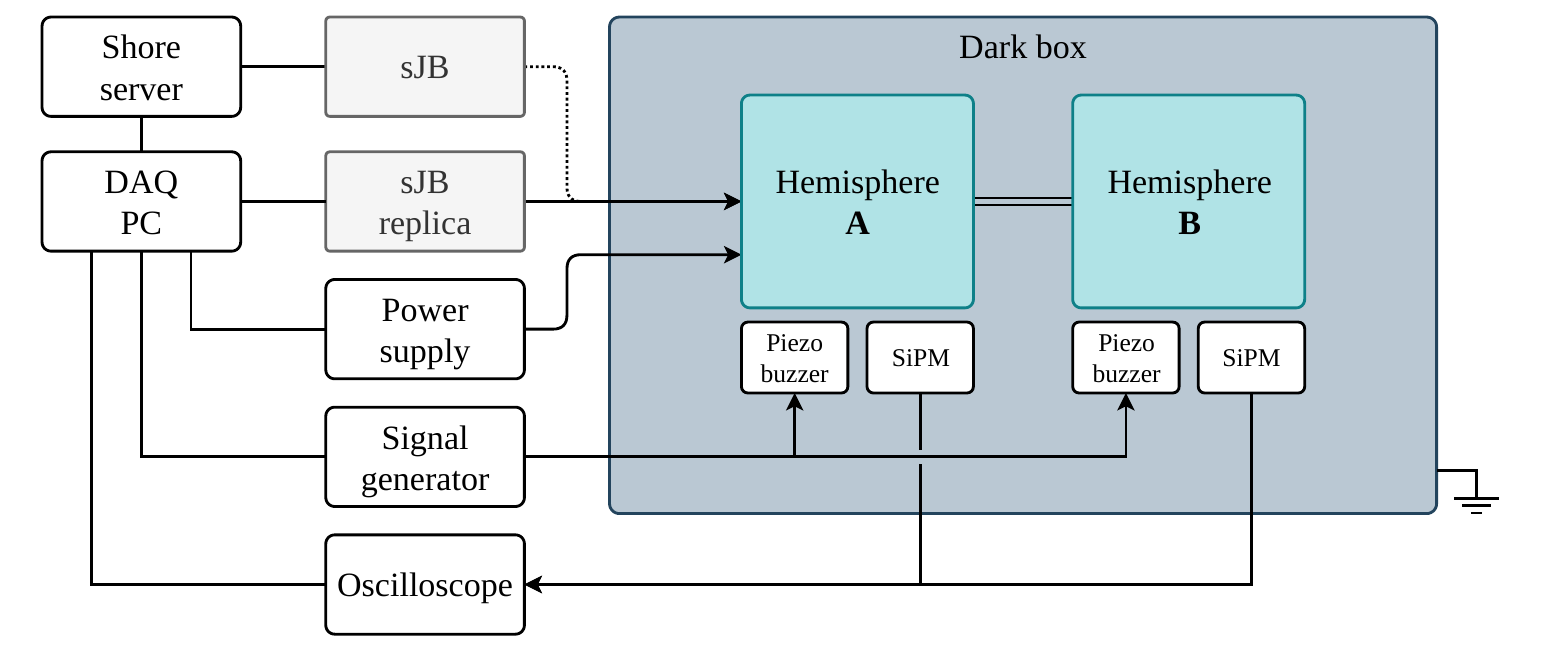}
    \caption{Experimental layout of the Module Integration and Network Test (MINT) system for final acceptance testing of module hemisphere pairs.}
    \label{fig:mint}
\end{figure}

The experimental layout of the MINT system is shown in \cref{fig:mint}. It consists of a two-door Faraday-cage dark box that enables the testing of two hemispheres. It provides a variety of electrical interfaces, including optical fibers for module network connectivity and timing, and replicates the sJB device architecture for as-built testing of the line. In addition to self-consistent tests of module functionalities through device read-backs, MINT incorporates external instruments for specific test cases. For one, piezo-acoustic receivers are validated using frequency signals from an external piezo-acoustic buzzer, mounted outside the glass hemisphere and driven by a function generator. In addition, optical flashers are validated with an external SiPM, using trigger counting on an external oscilloscope to assess light output. A dedicated computer is integrated into the network for MINT operation.
\par
The software functionality for MINT is implemented in the andromeda codebase, built around a dedicated backend testing infrastructure that executes standardized tests for each subsystem. This enables per-component customization of test configuration and sequencing, tailored validation of relevant data products and read-backs, and a unified software, logging, and data-taking interface across all components. The module test is encoded in a MIDAS sequence, which the user can configure and execute via either the web frontend hosted by the MIDAS server or the command line. An ordered sequence of tests ensures that each subsystem is initialized and verified before subsequent dependent tests can run. 

\begin{figure}[h!]
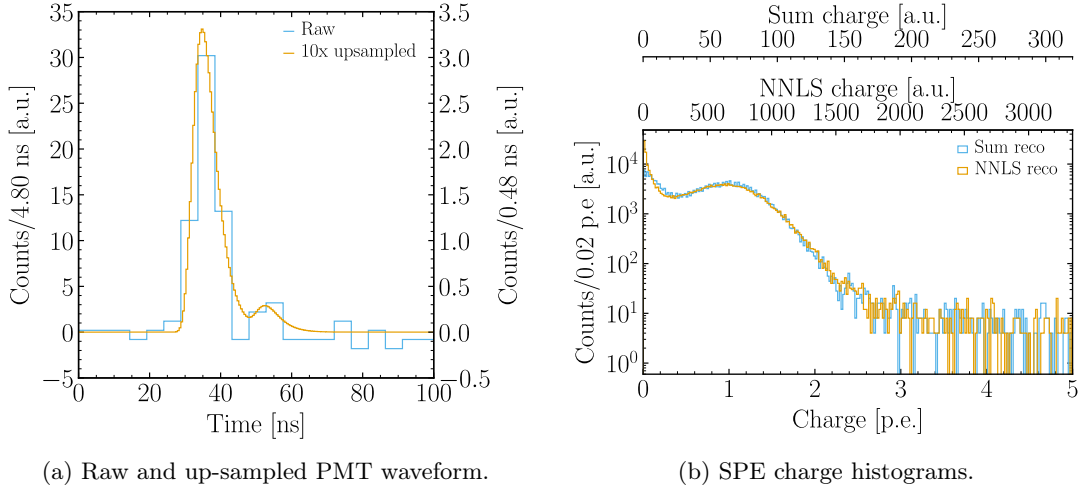

    \centering
    \begin{subfigure}[b]{0.46\textwidth}
        \centering
        \includesvg[width=\textwidth]{figs/plots/mint/upsampled_waveform} % Do not include .svg extension
        \caption{Raw and up-sampled PMT waveform.}
        \label{fig:mint-waveform}
    \end{subfigure}
    \quad
    \begin{subfigure}[b]{0.46\textwidth}
        \centering
        \includesvg[width=\textwidth]{figs/plots/mint/spe_spectrum} % Do not include .svg extension
        \caption{SPE charge histograms.}
        \label{fig:mint-spe-spectrum}
    \end{subfigure}
    \caption{PMT SPE response from MINT data. (a) Example waveform in its raw digitized format and up-sampled by a factor of $10$ using the NNLS deconvolution algorithm~\cite{daq_inprogress}. (b) SPE charge distribution obtained from the summation of raw-digitized and up-sampled waveforms.}
    \label{fig:mint-spe}
\end{figure}

In addition to validating the functionality of all systems, waveform data were recorded with both the PMTs and the acoustic receivers. In an offline step, a waveform analysis is performed to ensure that signal properties meet required specifications. For the PMTs, a non-negative least squares (NNLS) algorithm is used to up-sample the waveforms by a factor of 10, approximating the PMT waveform with a Gumbel distribution. A template fit is applied, which suppresses noise and improves the timing resolution to $<\,1\,$ns; see \cite{daq_inprogress} for a detailed description. \Cref{fig:mint-spe} shows a typical single photoelectron (SPE) waveform and spectrum obtained from dark-count measurements.

With this, the gain of each PMT was evaluated using MINT data. The nominal gain was initially set through the PMT operating voltage, based on measurements at TU Munich (TUM)~\cite{pmt_inprogress} where available, and otherwise on bulk measurements at the IceCube Upgrade test stand at RWTH Aachen~\cite{halveValidationOpticalSensors,IceCube:2024P07038}. These measurements were repeated in MINT on fully integrated modules, with the operational gain of each PMT set such that the SPE signal-to-noise ratio lies between $16$ and $31$, as defined for timing specifications. The gain response is linear over the entire range of interest. The operating voltages at nominal gain differ at the percent level between TUM and MINT data, which we attribute to MINT using the final integrated backend electronics and digitization scheme.

\begin{figure}[h!]
    \centering
    \includesvg[width=0.95\textwidth]{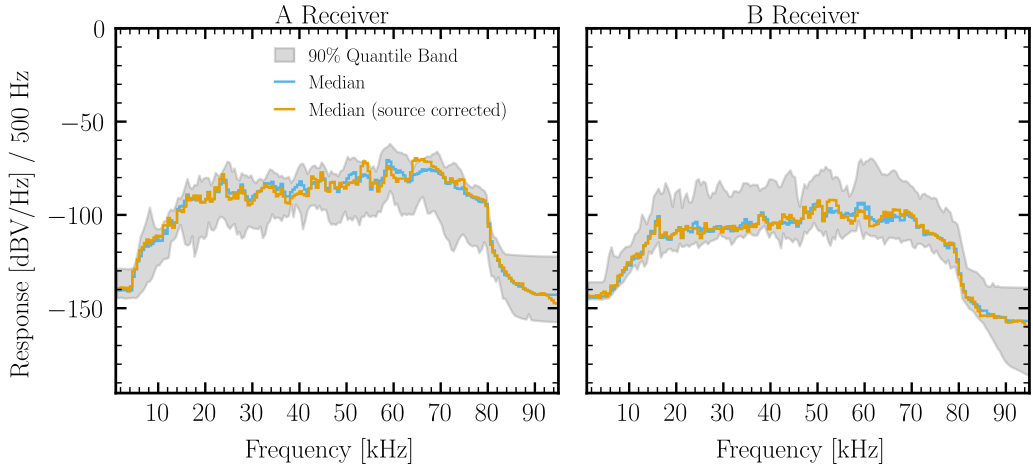}
    \caption{Frequency analysis of MINT data for all tested acoustic receivers showing the relative median frequency response with and without source-spectrum correction. Data taken at \unit[4]{dB} ADC channel gain and \unit[$2$]{$\text{V}_{\text{RMS}}$} full scale differential input.}
    \label{fig:mint-acoustic}
\end{figure}

The acoustic receivers, critical for positioning, perform remarkably consistently across the \unit[$10-60$]{kHz} frequency range relevant for positioning. Averaged over all measured sinusoidal frequencies and all modules, the A-side receivers reach a mean SNR of approximately \unit[45]{dB}, whereas the B-side receivers reach only approximately \unit[28]{dB}. This difference is attributed to the substantially weaker external B-side source. \Cref{fig:mint-acoustic} gives the median frequency response and \unit[90]{\%} quantile band across all modules for each receiver. Differences in acoustic coupling are expected due to the stochastic nature of the gluing process.
%\par
\par
A total of two PMTs and one acoustic receiver returned invalid waveform data during MINT testing of all modules; a failure rate of \unit[0.6]{\%} and \unit[2.5]{\%}, respectively. While an exchange of these components is possible, these were accepted to alleviate risk associated with disassembly and impact on the integration timeline.

%%%%%%%%%%%%%%%%%%%%%%%%%%%%%%%%%%%%%%%%%%%%%%%%
% Line integration
%%%%%%%%%%%%%%%%%%%%%%%%%%%%%%%%%%%%%%%%%%%%%%%%

\subsection{Line integration}
\label{subsec:line-integration}

Modules that passed acceptance testing are prepared for installation on the line. This involves visual inspection; cleaning of sealing surfaces of the hemisphere, the titanium cylinder, and the LC/UPC fiber connectors; installation of the sealing O-rings with lubricant \cite{Molykote111}; and integration and sealing of both hemispheres onto the titanium cylinder.

\begin{figure}[h!]
    \centering
    \begin{subfigure}[h]{0.43\textwidth}
        \centering
        \includegraphics[width=\textwidth]{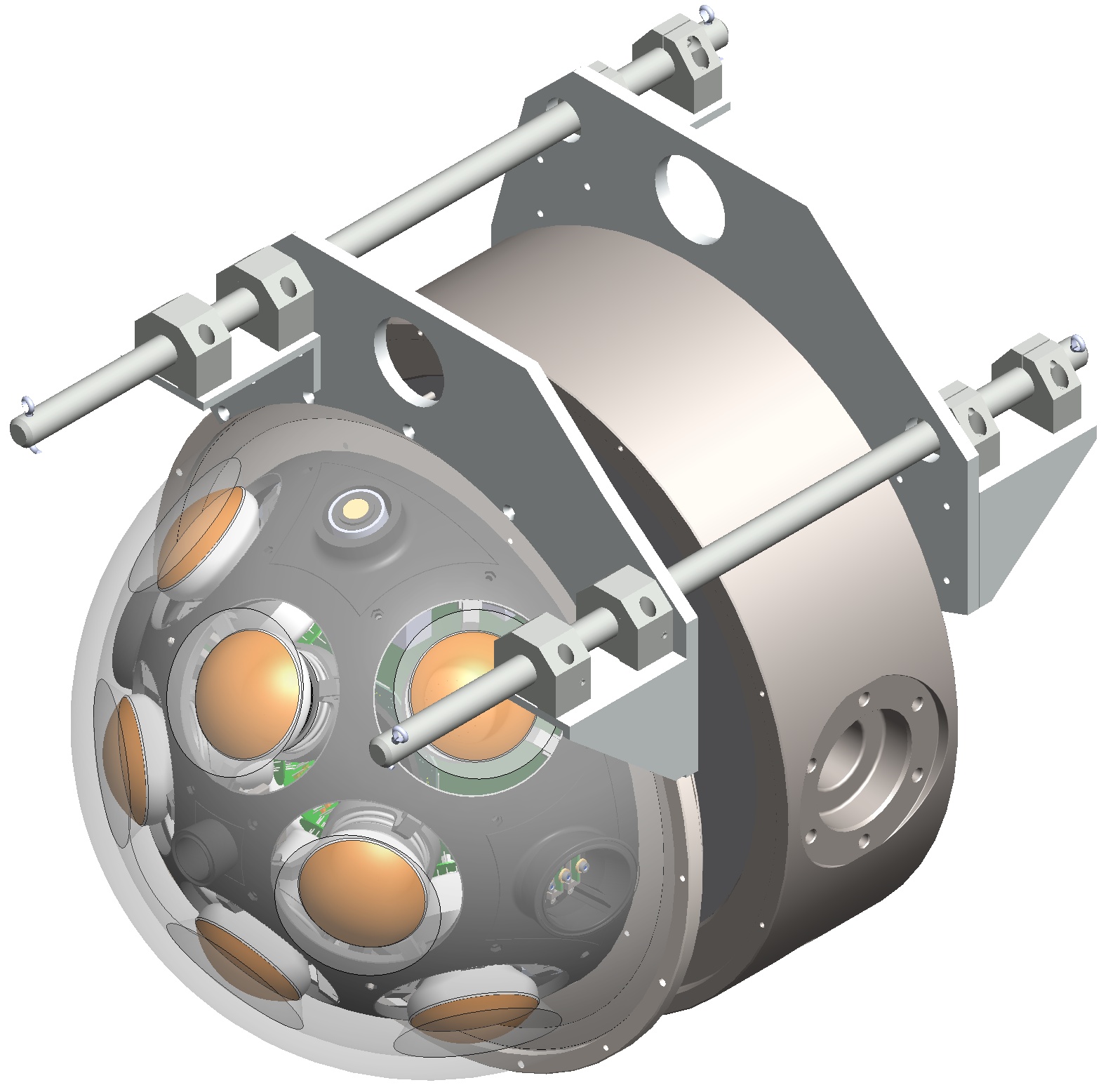}
        \caption{Integration jig rendering.}
        \label{fig:jig-a}
    \end{subfigure}
    \hfill
    \begin{subfigure}[h]{0.56\textwidth}
        \centering
        \includegraphics[width=\textwidth]{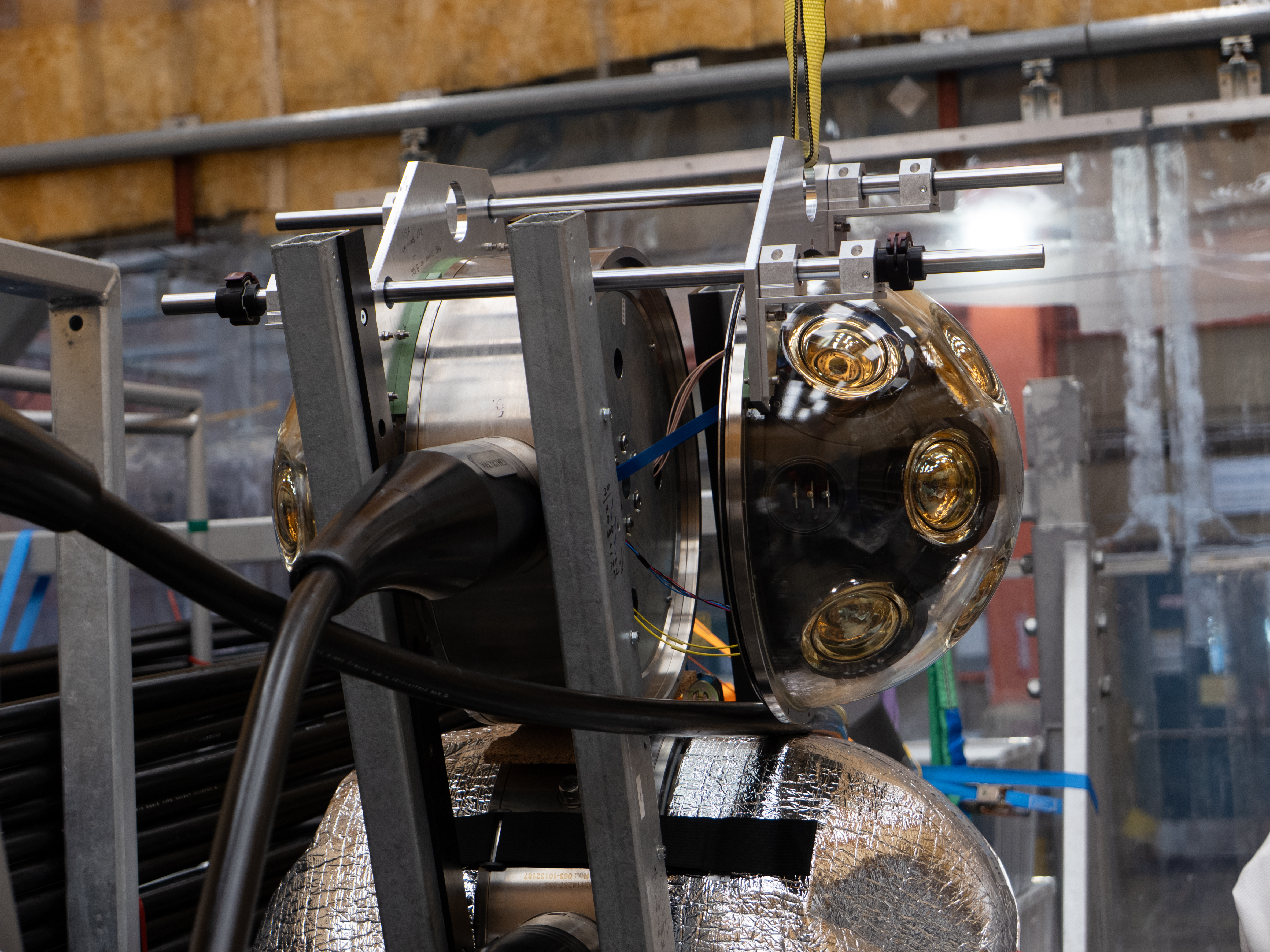}
        \caption{Integration jig at TRIUMF.}
        \label{fig:jig-b}
    \end{subfigure}
    \caption{(\subref{fig:jig-a}) Rendered view of the integration jig, showing the concept of aligning hemispheres with the titanium cylinder; adapted from~\cite{Spannfellner.2026}. (\subref{fig:jig-b}) Photograph of the hemisphere integration jig during assembly of an A-side hemisphere.}
    \label{fig:jig}
\end{figure}

The corresponding instrument termination on the detector line is prepared separately, using a jig specifically designed for module integration, as shown in \cref{fig:jig}. When affixed to the jig, the hemisphere can translate back and forth freely on four bearings, which enables cable routing and aligned integration of hemispheres.
\par
Modules are arranged in stacks of five on the deployment frame, with module number and type shown in \cref{fig:line}. Installation begins with the B-side hemisphere on the frame's inside: the hemisphere is connected to the jig, lifted into position, aligned with the titanium cylinder, and the PMT signal and fan-out board cables are fed through the cylinder's central opening before the hemisphere flange is bolted on. The jig is then reconfigured for the A-side hemisphere, which carries the mainboard. Here, the B-side cables, the main cable's power and serial connections, and the two timing and data fibers are connected, and the module's communication and response are verified before the A-side hemisphere is also bolted to the titanium cylinder. All fasteners are tightened per defined procedures.

To seat the O-rings, the assembled module is evacuated to approximately \unit[500]{mbar} and held for about one hour. The titanium bolts joining the cylinder and hemisphere flange are then reopened to add thread-locker (Loctite~271~\cite{Loctite.271}) for transport security. The module is subsequently dried through repeated cycles of evacuation down to \unit[300]{mbar} and nitrogen flushing through an integrated pressure valve, ending at an internal pressure of about \unit[900]{mbar}. Environmental sensor data (pressure, temperature, humidity) from inside the module is logged throughout this procedure, and whenever the module is powered.

After installation, module subsystems are tested over the detector line via the network and timing chain through the sJB. Acceptance requires full functionality of the mainboard, TOMcat, and fan-out board, plus compliance with PMT baseline noise and timing requirements. All twenty modules passed this test.

%%%%%%%%%%%%%%%%%%%%%%%%%%%%%%%%%%%%%%%%%%%%%%%%
% Frame integration
%%%%%%%%%%%%%%%%%%%%%%%%%%%%%%%%%%%%%%%%%%%%%%%%

\subsection{Frame integration}
\label{subsec:frame-integration}

Deployment frame finalization starts with the sJB. Before closing, desiccant pouches are added to remove residual moisture. The line cable terminates in a water-block, which forms the line-side closure of the sJB. The cable's conductors and fibers pass through the water-block chamber into the main sJB enclosure, where the conductors are wired to two DB-25 connectors and the fibers are individually fitted with LC/UPC connectors and routed into the optical network infrastructure. The water-block is filled with silicone oil to prevent water ingress into the sJB. Before the sJB is closed, a functional test verifies connectivity to each module. The node side is then sealed with a bolted titanium flange, which integrates the wet-mateable termination for the later connection to NEPTUNE.

The sJB rests horizontally in the subframe, with a mooring harness mounted to it to transfer the mechanical load during line unfurling. A bend restrictor protects the cable segment terminating at the water-block from over-bending as the sJB is lifted out of the subframe and rotated into the vertical. The subframe is mounted beneath the mainframe, and a spreader cage for lifting, placed on top of the mainframe, completes the assembly. The spreader cage is connected via shackles that a remotely operated vehicle (ROV) can actuate, allowing it to be removed subsea.

During storage, the light-sensitive PMTs and impact-sensitive glass housings are shielded by module covers, and a protective tarp with custom foam inserts guards the structure against mechanical impact.

%%%%%%%%%%%%%%%%%%%%%%%%%%%%%%%%%%%%%%%%%%%%%%%%
% System testing, commissioning
%%%%%%%%%%%%%%%%%%%%%%%%%%%%%%%%%%%%%%%%%%%%%%%%

\section{System-level validation}
\label{sec:line-testing}

We conducted several full-line functionality tests of the integrated system while preparing the line for mobilization and deployment: software upgrades and tests; functionality tests of the slow and fast data streams; performance tests of network, timing, and noise; and general monitoring of long-term operation. 

Testing began in air at TRIUMF, continued in water at Granville Island in Vancouver harbor, and concluded in air at ONC's Marine Technology Centre (MTC) on Vancouver Island, Canada -- the staging area for deployment.

%%%%%%%%%%%%%%%%%%%%%%%%%%%%%%%%%%%%%%%%%%%%%%%%
% Dry test
%%%%%%%%%%%%%%%%%%%%%%%%%%%%%%%%%%%%%%%%%%%%%%%%

\subsection{Line functionality}
\label{subsec:line-testing:dry}
% - Motivation: validate the full line as an integrated DAQ system before
%   water exposure.
% - Validation software framework: language/environment, key libraries,
%   test structure and configuration management, output and report format,
%   version control and traceability.
% - Test configuration: all modules connected via flight cables and SJB,
%   interfaced to the shore DAQ chain.
% - Network load tests: bandwidth utilisation, packet loss, and latency
%   under representative trigger rates.
% - DAQ system integration: timing synchronisation, run control, data
%   quality monitoring.
% - Results and any issues resolved prior to wet test.

\begin{figure}[h!]
    \centering
    \includesvg[width=\textwidth]{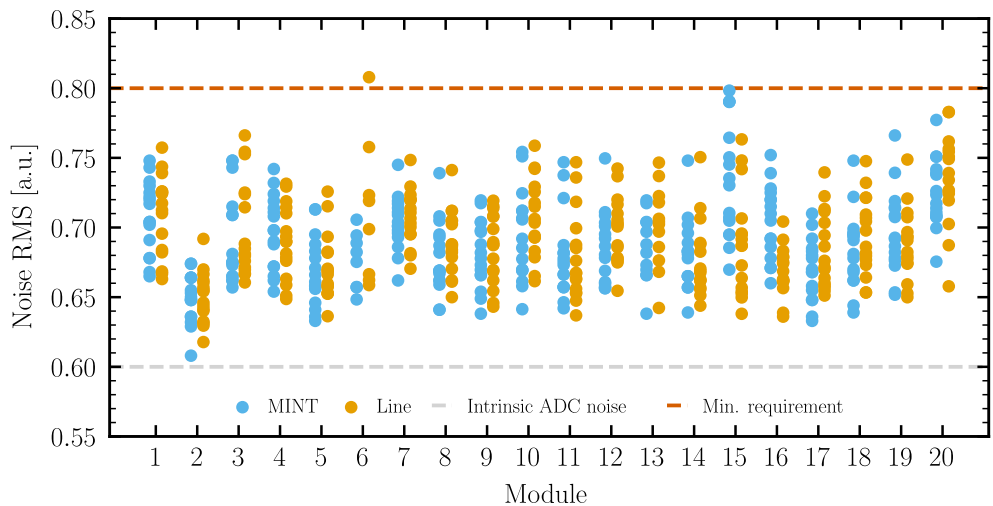} % Do not include .svg extension
    \caption{Baseline root-mean-square (RMS) values for all PMTs during MINT testing (blue) and after line integration (orange). The lower dashed line indicates the intrinsic ADC noise floor; the upper line marks the P-ONE-1 requirement.}
    \label{fig:line-vs-mint-noise}
\end{figure}

With all core systems active and initialized, modules consume \unit[$11.2 \pm 0.5$]{W} on average, with a supply voltage of \unit[96]{V}. Because each module is powered through an individual conductor pair, a varying number of energized conductor wires run past each module within the titanium can, increasing the risk of induced noise. We find that our noise mitigation compensates for this effect, and that the baseline noise level is independent of the module position. As shown in \cref{fig:line-vs-mint-noise}, comparable noise profiles are measured in all modules on the fully powered line, with up to $20$ energized conductors passing a module.
\newpage
The second key test is saturating the network bandwidth with all data streams to understand potential performance limitations. The combined streams of slowDAQ, acoustic waveforms, and camera images only consume about \unit[10]{\%} of the available \unit[1]{Gbps} bandwidth, which means stress testing requires operating the fastDAQ, ideally without operating PMTs to avoid accidental light exposure. We achieve this by configuring the fastDAQ handlers on all modules to send sporadic artificial triggers at a defined average rate. The shore handler collates the trigger packets from all modules. It executes waveform readouts accordingly, so that every module receives waveform readout requests at the same frequency while the shore handles 20 times the per-module trigger rate.

Several network stress tests were performed on the integrated system, both at TRIUMF and at MTC. The tests ranged from a few minutes to several hours and targeted the identification of network issues in preparation for deployment. Packet monitoring at all network junctions allows traffic to be traced and bottlenecks to be localized. For example, severe packet loss observed after line integration at TRIUMF was traced back to a faulty Ethernet cable in the testing backend infrastructure. Network performance has been stable and predictable since. During a representative $2$-hour fastDAQ run at \unit[75]{kHz} of forced triggers, we recorded $1.1\times10^{10}$ successful line readouts on shore, corresponding to about $1.7\times10^{11}$ acquired (artificial) PMT waveforms.

%%%%%%%%%%%%%%%%%%%%%%%%%%%%%%%%%%%%%%%%%%%%%%%%
% Wet test
%%%%%%%%%%%%%%%%%%%%%%%%%%%%%%%%%%%%%%%%%%%%%%%%

\subsection{Submersion test}
\label{subsec:line-testing:wet}
% - Test configuration: full line or representative subset submerged in a
%   controlled water volume at TRIUMF.
% - Ground fault isolation: verification of electrical safety and insulation
%   integrity in water.
% - Optical performance: PMT dark rate, inter-module coincidence rate,
%   flasher visibility between modules.
% - Acoustic performance in water if measured.
% - Comparison to dry-test and MINT baselines; confirmation of no degradation.

After completing integration, the line was mobilized to Granville Island Boatyard\footnote{Granville Island Boatyard, Canada. \url{https://www.granvilleislandboatyard.com/}} in Vancouver, Canada, for a submersion test, also referred to as the "wet test". The purpose of this test was to confirm in-water operation of the detector, verify the grounding connection to seawater, and conduct software tests. On-site activities were led jointly by P-ONE and ONC teams.

\begin{figure}[h!]
    \centering
    \begin{subfigure}[b]{0.4\textwidth}
        \centering
        \includegraphics[width=\textwidth]{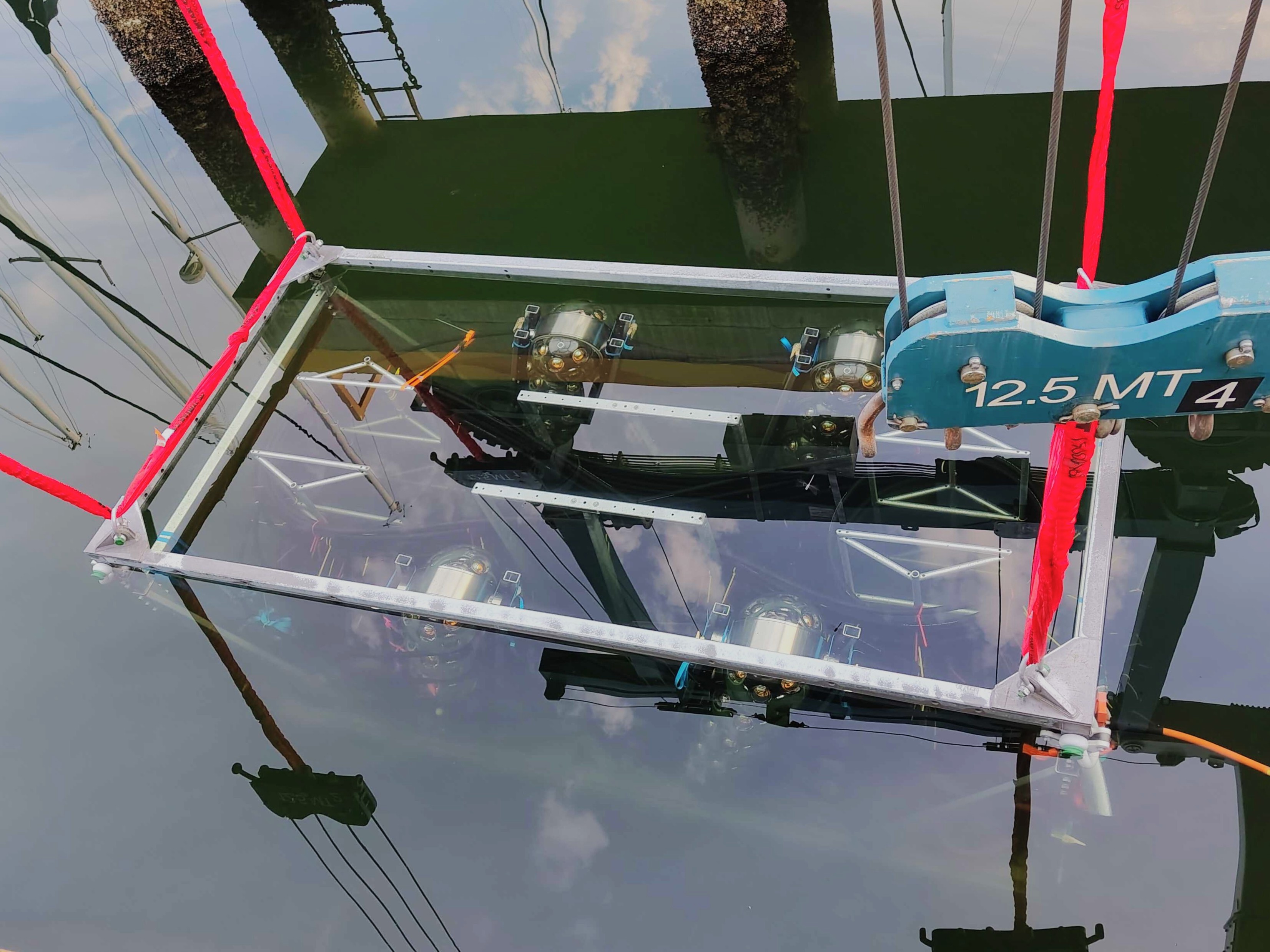}
        \caption{Submerged line.}
        \label{fig:wet-test-line}
    \end{subfigure}
    \hspace*{1.5cm}
    \begin{subfigure}[b]{0.3\textwidth}
        \centering
        \includegraphics[width=\textwidth]{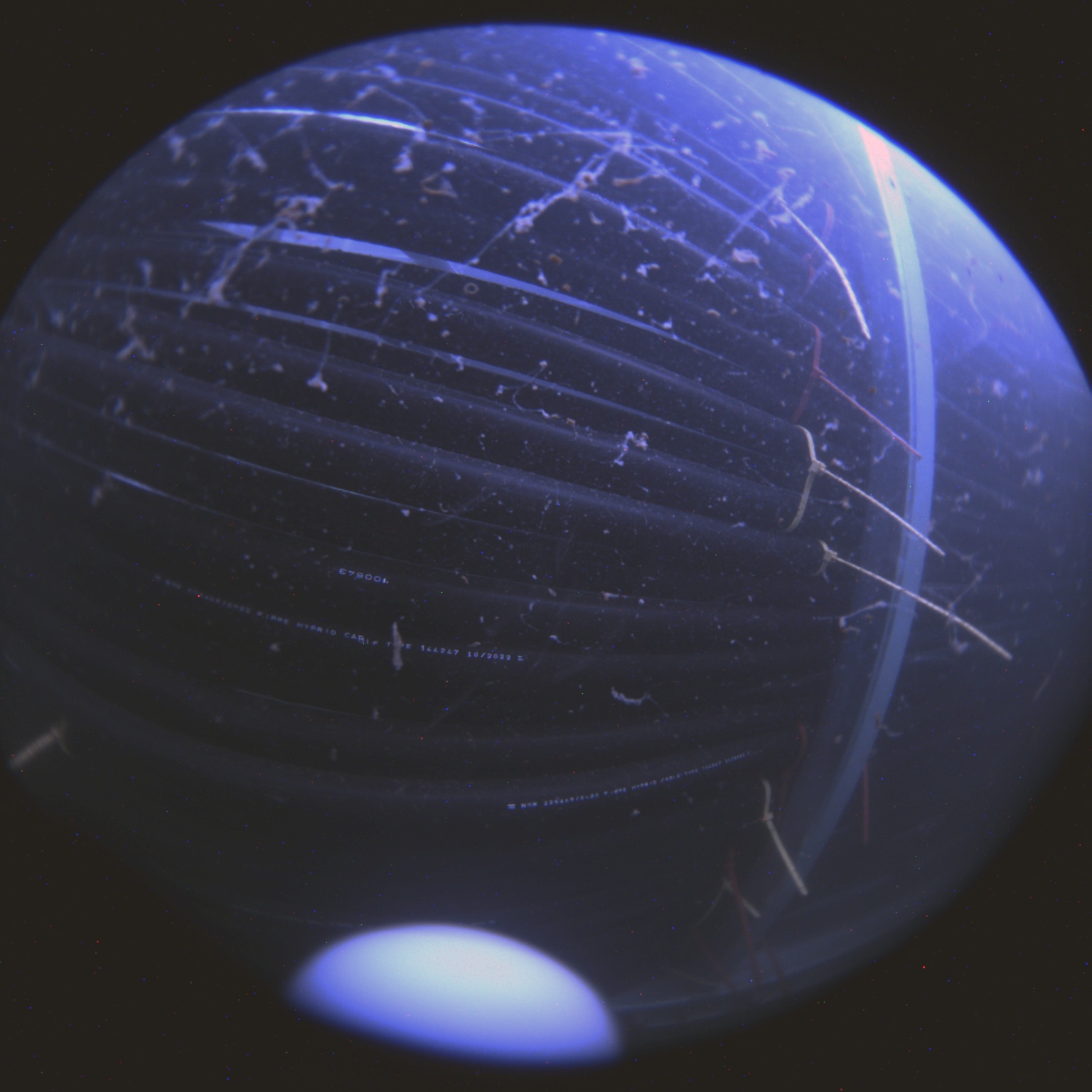}
        \caption{P-CAL camera view.}
        \label{fig:wet-test-pcal-b}
    \end{subfigure}
    \begin{subfigure}[b]{0.99\textwidth}
        \centering
        \includegraphics[width=\textwidth]{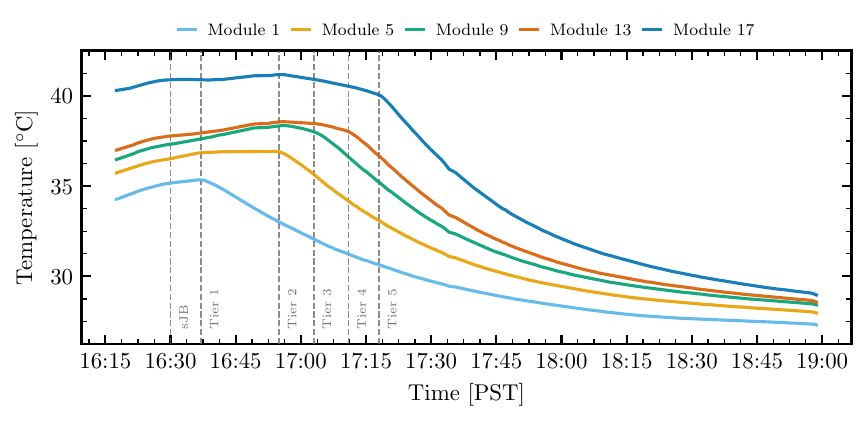}
        \caption{Module stack temperature profiles.}
        \label{fig:wet-test-pth}
    \end{subfigure}
    \caption{Wet test of the P-ONE-1 line at Granville Island boatyard. (a) The line after full submersion. (b) Camera view of a B-side P-CAL hemisphere after submersion, looking into the cable stack, and (c) Internal temperature of a module stack during the submersion process. Vertical lines mark water contact of respective tiers and the expected, staggered cooling effect is visible.}
    \label{fig:wet-test}
\end{figure}

The line was submerged using a movable boat crane after mechanical lifting preparations to the frame -- \cref{fig:wet-test-line}. The actual submersion started with the sJB and was performed in stages, submerging instrument tiers one after the other. At each tier, submersion was paused for about \unit[10]{minutes} to monitor instrument telemetry and the grounding connection to seawater for short-circuit faults, which would suggest water ingress. After approximately one hour, the line was fully submerged without observed electrical faults or water ingress. P-CAL camera images and temperature readings for a module stack during submersion are shown in \cref{fig:wet-test-pcal-b,fig:wet-test-pth}, respectively. Each module tier shows a temperature decrease that coincides with its submersion, demonstrating the expected cooling effect of the surrounding water.

Several service, test, and monitoring tasks were completed during the 14-hour test, including firmware upgrades to critical infrastructure, acoustic and camera data taking, and operational software testing in preparation for deployment. 
\par
%

% %%%%%%%%%%%%%%%%%%%%%%%%%%%%%%%%%%%%%%%%%%%%%%%%
% MTC commissioning
% %%%%%%%%%%%%%%%%%%%%%%%%%%%%%%%%%%%%%%%%%%%%%%%%

\subsection{Deployment preparation}
\label{subsec:line-testing:mtc}
% - Introduce the MTC site and its role: shore station operated by ONC,
%   hosting the shore computing infrastructure and cable landing for P-ONE-1.
% - Summarise activities performed at MTC: extended operational software
%   testing, final commissioning of the shore computing infrastructure, and
%   commissioning of the acoustic seafloor positioning system.
% - State the outcome: software and infrastructure met commissioning
%   criteria; line and positioning system verified ready for deployment.
% - Note any issues identified and resolved during commissioning, and their
%   implications for deployment readiness.
% - Transition into deployment preparations following MTC commissioning.

After the wet test, the line was mobilized to MTC. On-site work in preparation for deployment included further testing of the operational software, commissioning the shore computing infrastructure and the acoustic seafloor positioning system, and mechanical finalization of the frame.

To enable continuous operation of the line at MTC, an external water-cooling system was built and installed to cool the sJB while operating in air. With this in place, the line ran almost continuously for roughly four weeks, with an on-site team leading operation.

% %%%%%%%%%%%%%%%%%%%%%%%%%%%%%%%%%%%%%%%%%%%%%%%%
% Operational software
% %%%%%%%%%%%%%%%%%%%%%%%%%%%%%%%%%%%%%%%%%%%%%%%%

\subsubsection{Operational software}
\label{subsubsec:ops-software}

A core objective was to establish a stable software state for deployment, particularly for unfurling. Because an ROV actively monitors the line during startup and unfurling, the associated procedures must be reliable and fast. This includes starting up the line and enabling the slowDAQ and acoustic data streams to monitor the unfurling process.

Startup, shutdown, and data-taking procedures were tested on the existing codebase, comprising MIDAS, andromeda, and perseus. Testing also drove the development of missing control code and patching of unexpected behavior encountered along the way. The MIDAS control code and web frontends for detector operation, acoustic and camera control, and the interface to ONC's data portal Oceans3.0 were finalized in the process. On site, the line was fully rebooted about twice per day, and several software patches were identified and deployed to improve the reliability of automated procedures; after optimization, a full reboot takes about \unit[15]{minutes}. During on-site operations, we achieved a slowDAQ uptime of \unit[88.6 (57.1)]{\%} excluding (including) planned downtime.

\begin{figure}[h!]
    \centering
    \includegraphics[width=0.99\textwidth]{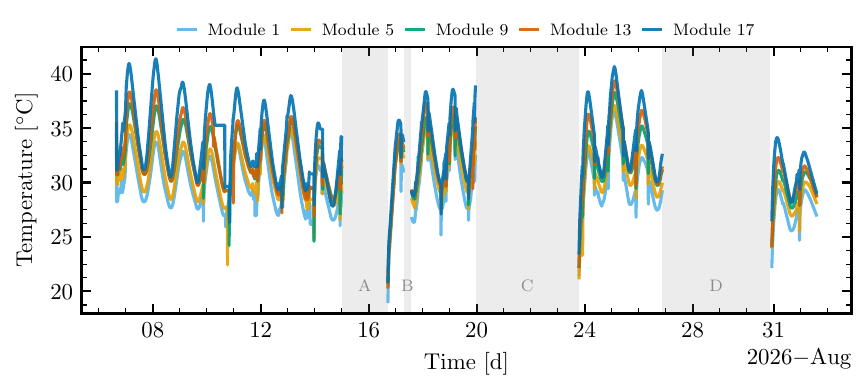}
    \caption{Monitored temperature data for a module stack at MTC, collected via MIDAS using the slowDAQ handler and down-sampled. Downtimes are associated with operations requiring full power-down (A), active slowDAQ development (B), or periods with limited line access (C, D).}
    \label{fig:slowdaq-mtc}
\end{figure}

Temperature monitoring of a module stack is shown in \cref{fig:slowdaq-mtc}. Day-night temperature variations are clearly visible, and downtime is associated either with scheduled power-downs on weekends when site access was limited or with major line operations requiring shutdown of the slowDAQ.
\par
%

% %%%%%%%%%%%%%%%%%%%%%%%%%%%%%%%%%%%%%%%%%%%%%%%%
% Acoustic positioning system
% %%%%%%%%%%%%%%%%%%%%%%%%%%%%%%%%%%%%%%%%%%%%%%%%

\isolate{
\subsubsection{Acoustic positioning system}

The on-site activities also included commissioning of the acoustic seafloor positioning system procured from Sonardyne Ltd\footnote{Sonardyne Ltd., United Kingdom. \url{https://www.sonardyne.com}}. The system comprises three battery-powered acoustic pingers (FETCHs~\cite{SonardyneFetch.2025}) and a cabled transceiver (Dunker 6~\cite{,SonardyneDunker6+.2025}) connected to the sJB. All devices are equipped with omnidirectional piezo-acoustic elements with uniform angular pressure profiles in water, and can achieve a transmit source level of up to \unit[196]{dB re $1\,\mu$Pa at $1\,$m}. A similar system has been ocean-tested using a P-ONE prototype module~\cite{AcousticCal.2025}. Individual tripods will sit each device approximately \unit[$1-2$]{m} above the seafloor.

The transceiver is cabled to the sJB, which supplies the operating voltage, an RS-485 serial connection, and a trigger line over copper conductors. Serial communication is handled by the MIDAS, andromeda, and perseus software stack via dedicated control handlers and custom communication protocols (\cref{fig:code-stack}). The sJB's acoustic trigger system is integrated into the detector timing infrastructure and timestamps every interrogation cycle. Upon receipt of a trigger, the transceiver emits an interrogation pulse encoded with Sonardyne's Wideband\textsuperscript{\textregistered}~\cite{poole2007wideband} acoustic payload, which the pingers decode and respond to acoustically. A single interrogation can therefore address one or several units, query complex information, or carry new configurations for remote programming of the autonomous pingers. Each positioning cycle thus presents the piezo-acoustic receivers in all P-ONE modules with four acoustic signatures -- one interrogation and three responses -- from which the position of individual modules can be determined by multilateration.

At MTC, the Sonardyne system was commissioned, and its control software was tested and verified. All acoustic functionality was checked in air at short transceiver-to-pinger distances using the setup shown in \cref{fig:sonardyne}. The absolute positioning system (Sonardyne Ranger 2~\cite{sonardyne_ranger2}) was likewise commissioned and will be operated from the deployment vessel to position all devices in the GPS reference frame from the ocean surface.

Beyond direct acoustic interrogation between a transceiver and a pinger, the system can also relay information from one unit to the next, propagating commands across successive acoustic links. Relaying makes it possible to range directly between two autonomous units, rather than only between transceiver and pinger: a command sent to the first unit triggers an interrogation of the second, and the resulting two-way travel time yields the distance between them. This capability was also verified in air, and it will allow the inter-unit distances -- and hence the relative geometry of the seafloor units -- to be constrained in situ once deployed.
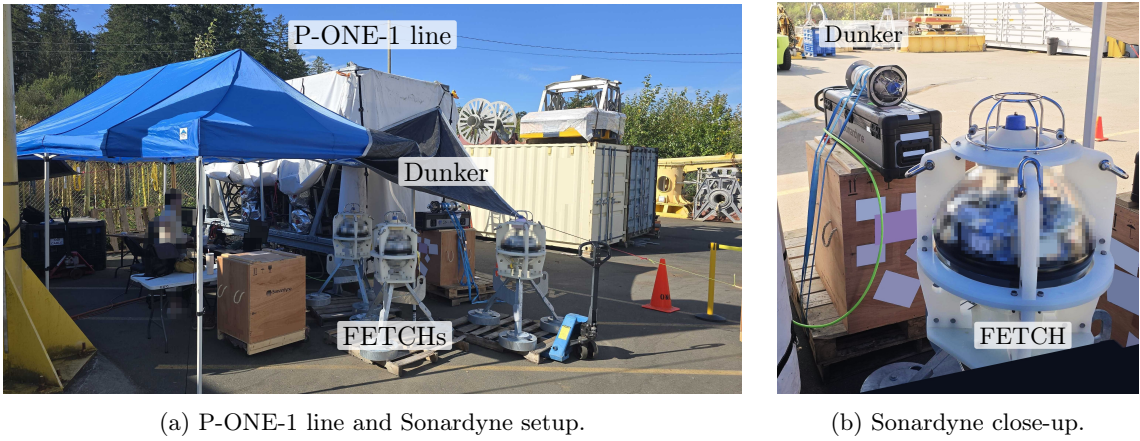
\begin{figure}[h!]
    \centering
    \tikzset{lbl/.style={fill=white, fill opacity=0.75, text opacity=1,
                         inner sep=2pt, rounded corners=1pt, font=\small}}
    \begin{subfigure}[b]{0.65\textwidth}
        \begin{tikzpicture}
            \node[anchor=south west, inner sep=0] (img) at (0,0)
                {\includegraphics[width=\linewidth]{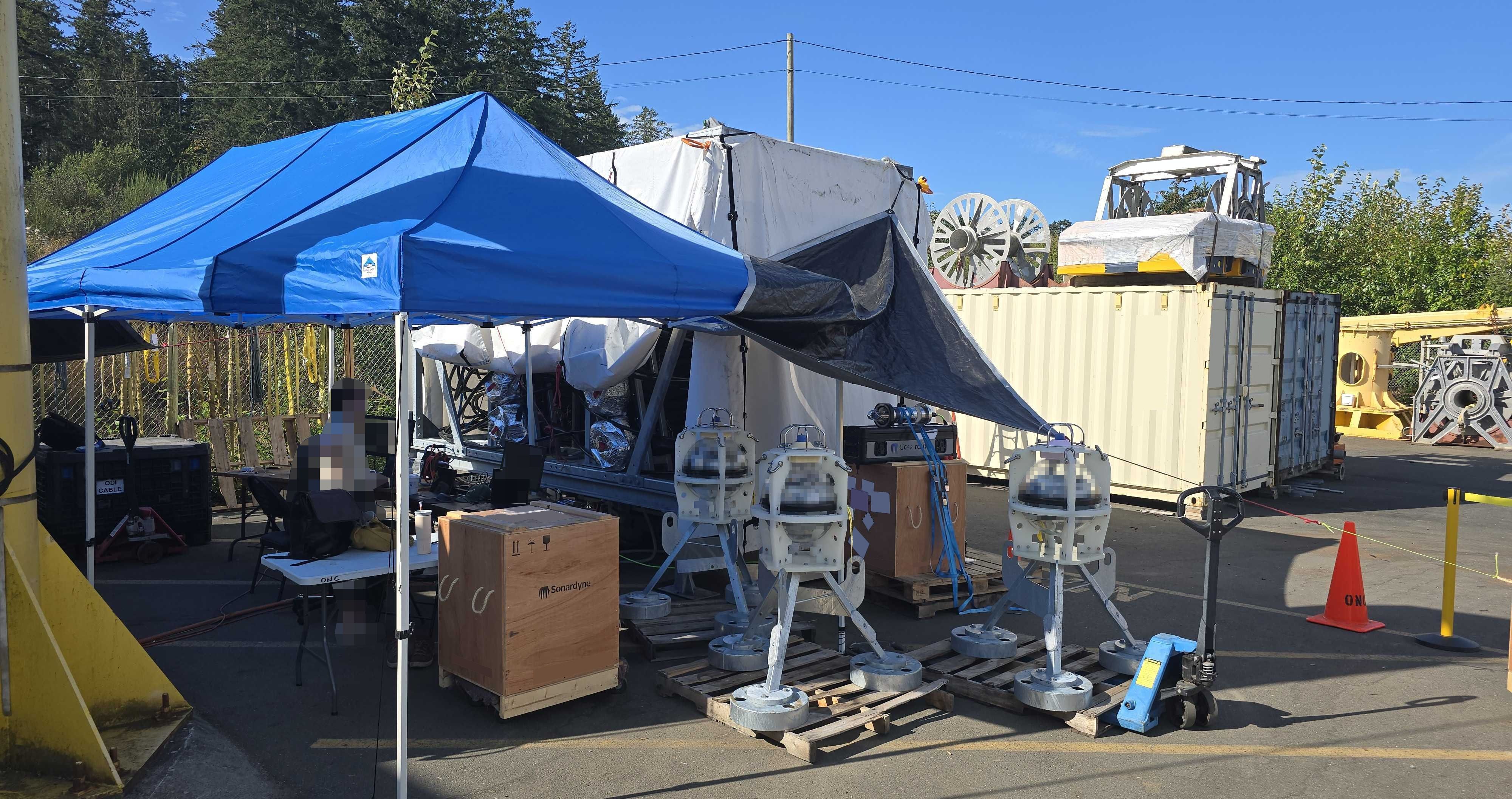}};
            % (0,0) = bottom-left of image, (1,1) = top-right
            \begin{scope}[x={(img.south east)}, y={(img.north west)}]
                % --- calibration grid: uncomment to read off coordinates ---
                % \draw[help lines,xstep=.1,ystep=.1] (0,0) grid (1,1);
                % \foreach \i in {0,...,9} \foreach \j in {0,...,9}
                %     \node[anchor=south west,font=\tiny,text=red] at (\i/10,\j/10) {\i\j};
                \node[lbl] at (0.50,0.92) {P-ONE-1 line};
                \node[lbl] at (0.53,0.15) {FETCHs};
                \node[lbl] at (0.60,0.57) {Dunker};
            \end{scope}
        \end{tikzpicture}
        \caption{P-ONE-1 line and Sonardyne setup.}
        \label{fig:sonardyne_a}
    \end{subfigure}
    \hfill
    \begin{subfigure}[b]{0.318\textwidth}
        \begin{tikzpicture}
            \node[anchor=south west, inner sep=0] (img) at (0,0)
                {\includegraphics[width=\linewidth]{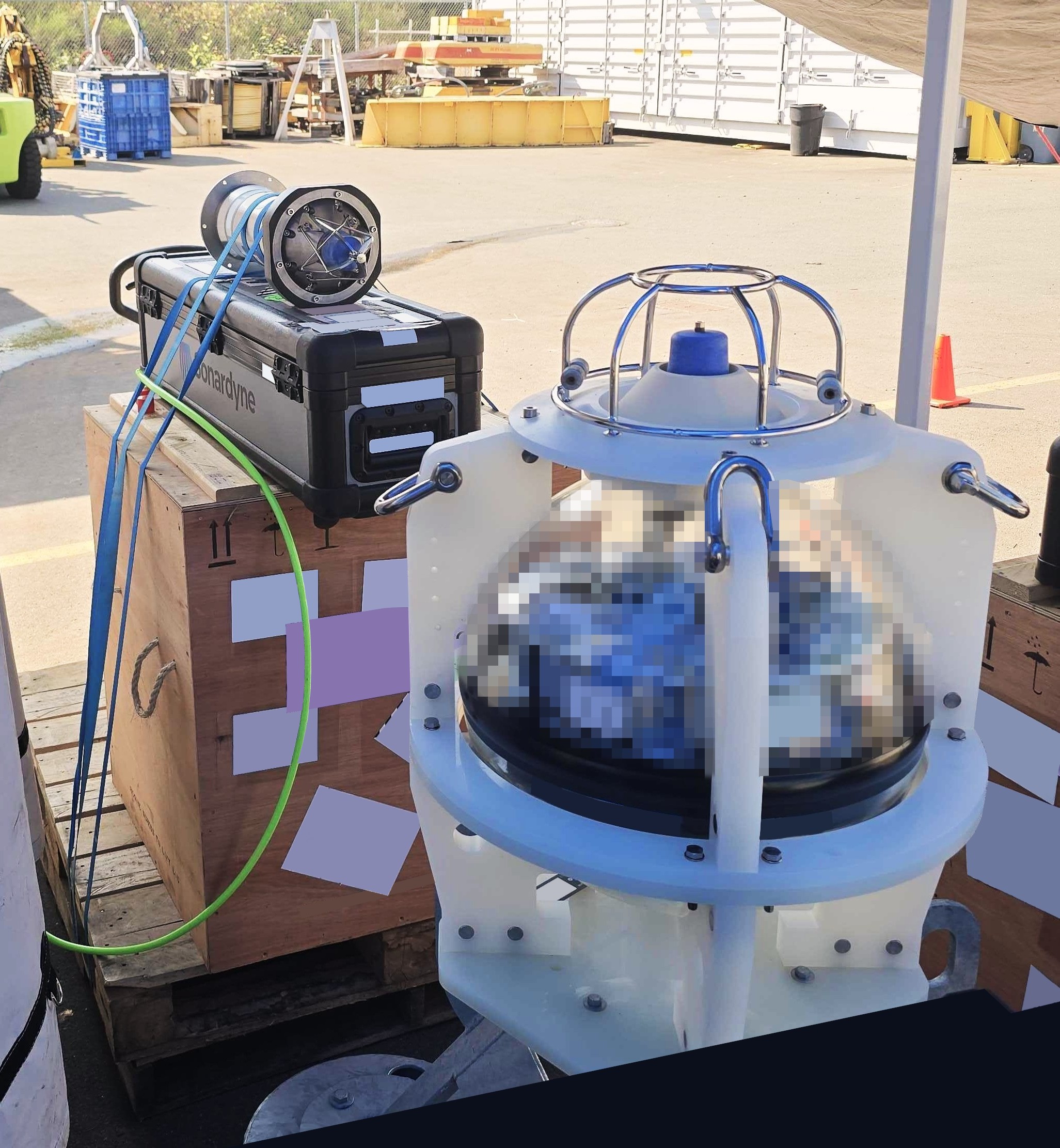}};
            \begin{scope}[x={(img.south east)}, y={(img.north west)}]
                % \draw[help lines,xstep=.1,ystep=.1] (0,0) grid (1,1);
                % \foreach \i in {0,...,9} \foreach \j in {0,...,9}
                %     \node[anchor=south west,font=\tiny,text=red] at (\i/10,\j/10) {\i\j};
                \node[lbl, font=\footnotesize] at (0.68,0.15) {FETCH};
                \node[lbl, font=\footnotesize] at (0.24,0.92) {Dunker};
            \end{scope}
        \end{tikzpicture}
        \caption{Sonardyne close-up.}
        \label{fig:sonardyne_b}
    \end{subfigure}
    \caption{P-ONE-1 and Sonardyne system setup at MTC.
    (a) P-ONE-1 covered in the background, with the autonomous pingers
    (FETCHs) and the cabled transceiver (Dunker) in the work area.
    (b) Close-up view of FETCH and Dunker; the Dunker connects to the sJB
    by cable.}
    \label{fig:sonardyne}
\end{figure}%\par
}

% %%%%%%%%%%%%%%%%%%%%%%%%%%%%%%%%%%%%%%%%%%%%%%%%
% Frame finalization
% %%%%%%%%%%%%%%%%%%%%%%%%%%%%%%%%%%%%%%%%%%%%%%%%

\subsubsection{Mechanical preparation}

\isolate{The mechanical preparation of the line is the final step before mobilization for sea operations. It comprises} visual inspection and preparation of the cable stack; preparation of the spreader cage, the subsea float, and its release mechanism for unfurling; rigging and fastening of the sJB within the subframe; \isolate{and mounting of the wet-mate connector panel and acoustic transceiver to the mainframe.}

On both the subframe and the mainframe, the cable layering is visually inspected, and individual cable segments are secured to the module stands to ensure controlled, incremental unfurling subsea. Cable ties added solely for transport safety are removed offshore immediately before the heavy-lift operation. The float release mechanism is mounted to the subframe and actuated by the ROV subsea to initiate unfurling. It secures the subsea float -- attached to the line on the back deck of the deployment vessel -- to the subframe until release. The float connects to the topmost module via a \unit[10]{m} synthetic rope and a titanium swivel, which decouples rotational forces on the float from the detector line. The sJB is rigged to the subframe with a Seaflex\textsuperscript{\textregistered}~\cite{Seaflex.2023} shock absorber and two safety wires. \isolate{In preparation for ROV operations, the wet-mate connector panel for the sJB/NEPTUNE connection and the acoustic transceiver are mechanically secured to the mainframe.}

% %%%%%%%%%%%%%%%%%%%%%%%%%%%%%%%%%%%%%%%%%%%%%%%%
% Subsea installation plans
% %%%%%%%%%%%%%%%%%%%%%%%%%%%%%%%%%%%%%%%%%%%%%%%%

\subsubsection{Subsea installation plans} \label{subsec:subsea_install}

P-ONE-1 will be deployed in proximity to the Cascadia Basin node of NEPTUNE as shown in \cref{fig:map}. The site offers favorable conditions for a neutrino detector: a water depth of \unit[2.6]{km}, a flat seafloor bathymetry with soft sediment, and moderate water currents of up to \unit[10]{cm/s}~\cite{Holzapfel.2023}. It is also free of major infrastructure, leaving space for future lines.

The Cascadia Basin node of NEPTUNE lies within the marine protected area (MPA) \ThT{}~(TḥT), established by the Nuu-chah-nulth Tribal Council, the Council of the Haida Nation, Pacheedaht First Nation, the Quatsino First Nation, and the Government of Canada~\cite{DFO.THT,ThT.2024}. The MPA protects unique seafloor features and their associated ecosystems, and therefore requires P-ONE to be installed outside its boundary. A suitable site for the P-ONE detector was identified about \unit[35]{km} east of the Cascadia Basin node, outside the MPA boundary. The connection between NEPTUNE and P-ONE is made with a custom extension assembly, comprising a cable of this length with voltage step-up and step-down converters at either end. At a later stage, the extension assembly will be reconnected to a P-ONE cluster junction box that will host P-ONE-1 together with additional lines, allowing the detector to be extended without immediate modification of NEPTUNE infrastructure. \isolate{The extension cable will be deployed and connected before P-ONE-1, together with the autonomous acoustic pinger units. The pingers will be released in free fall from the deployment vessel at their planned locations, and their absolute positions will be determined with a surface-based acoustic positioning system.}
\isolate{
\begin{figure}[h!]
    \centering
    \includegraphics[width=0.99\textwidth]{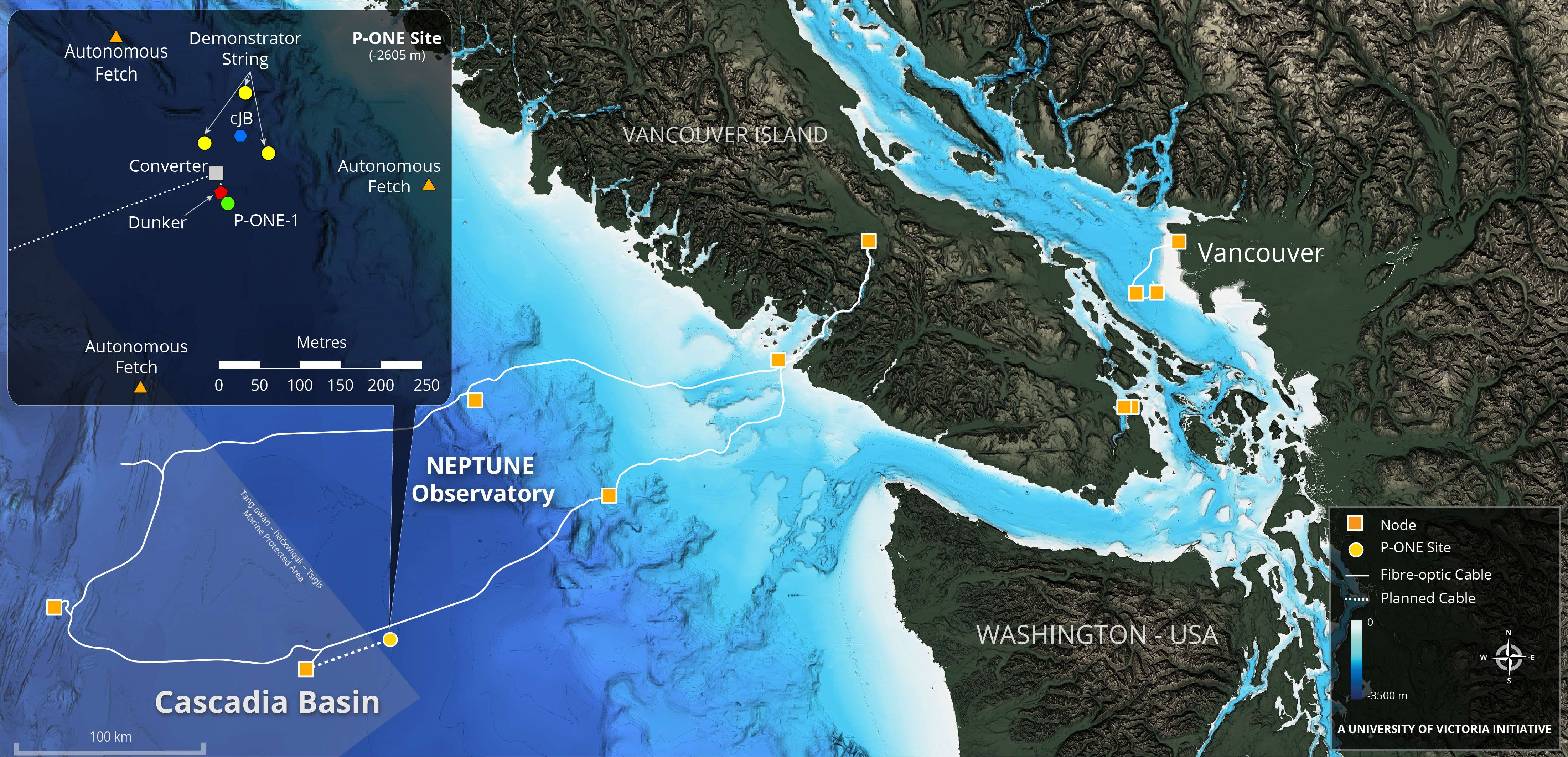}
    \caption{Location and subsea layout of P-ONE. The main map displays ONC's NEPTUNE cabled observatory infrastructure off Vancouver Island. The inset highlights the planned P-ONE-1 and P-ONE-D site geometry, including autonomous and cabled acoustic transponders, interconnecting cables, and the junction box layout. Map courtesy of Ocean Networks Canada.}
    \label{fig:map}
\end{figure}\par}

\begin{figure}[h!]
    \centering
    \includegraphics[width=0.99\textwidth]{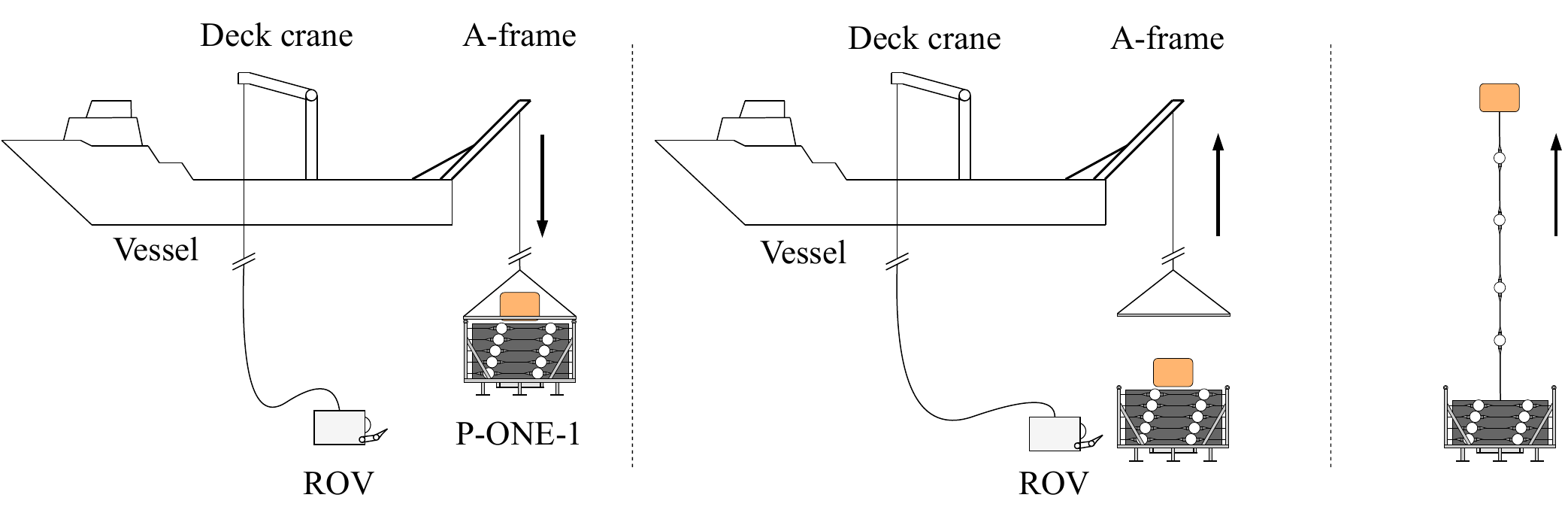}
    \caption{Deployment sequence of the P-ONE-1 line: heavy-lift lowering of the line (left), ROV removal of the spreader cage and line connection (center), and unfurling after actuation of the float release (right). Figure taken from~\cite{Spannfellner.2026}.}
    \label{fig:deployment}
\end{figure}

Sea operations for P-ONE-1 are planned during daylight hours. The line is lowered to the seafloor via the heavy-lift wire, with direct ROV support to monitor its touchdown. After touchdown, the ROV surveys the line visually and connects the sJB to the NEPTUNE infrastructure via a wet-mate cable. Clearance to proceed is given once visual inspection and several tiers of operational checks confirm detector line integrity. Unfurling preparations begin with two parallel activities: powering up the line and removing the spreader cage. These are followed by actuation of the float release and unfurling of the line. A sketch of these steps is shown in \cref{fig:deployment}.
\par
\isolate{After unfurling, the ROV detaches the acoustic transceiver from the line and moves it to its final position about \unit[30]{m} away.} It then returns to the line and ascends slowly, inspecting the cable and each module on the way back to the surface. Afterwards, the line enters commissioning in preparation for data-taking objectives.

%%%%%%%%%%%%%%%%%%%%%%%%%%%%%%%%%%%%%%%%%%%%%%%%
% Conclusion
%%%%%%%%%%%%%%%%%%%%%%%%%%%%%%%%%%%%%%%%%%%%%%%%

\section{Conclusion}
% - Summarise the commissioning campaign: design-driven distributed production,
%   integration at TRIUMF, and the full qualification chain from prototype
%   through system-level validation.
% - State the outcome: all 20 P-OMs and all line subsystems passed acceptance
%   criteria; P-ONE-1 is cleared for ocean deployment.
% - Anticipated deployment timeline and first-data outlook.
% - Lessons feeding into full array production planning.
This work presents the design and integration of the first P-ONE detector line, P-ONE-1, in preparation for its deployment to the Pacific Ocean. It details the functional principle of the P-ONE-1 design, covering both the line itself and the wider P-ONE detector architecture, as well as the integration and testing carried out at several sites across Canada.

The P-ONE-1 integration sequence started at TRIUMF with prototype hemisphere measurements in preparation for hemisphere integration and final acceptance testing in the MINT system. These measurements identified necessary mitigations for both the electrical noise on the PMT signal cables and the mechanical mounting infrastructure of the mainboard. With both in place, the noise performance and the vibration- and shock-test results meet or exceed the required specifications. In a second step, all modules were integrated into the line, and system-level tests were prepared and executed. All modules on the line continue to meet the noise requirements, even with up to 20 energized copper conductors in the vicinity of a module. Subsequent network tests confirmed the anticipated bandwidth limit of \unit[1]{Gbps} with the entire line participating in data-taking.

Two additional sites were subsequently used for line validation tests: a water submersion test at the Granville Island boatyard in Vancouver, and final line preparation at MTC, operated by ONC on Vancouver Island. The former was a 14-hour test with the entire line submerged in ocean water, assessing pressure housing integrity and the seawater grounding scheme and executing several critical software tests. It was passed without any observed issues. \isolate{At MTC, the line was powered for the majority of four weeks} with an uptime of about \unit[86.9]{\%} (not including planned power-downs) using a water-cooling system for the sJB \isolate{to finalize software readiness for deployment, commission the acoustic seafloor system, and complete the mechanical preparations of the line. All three objectives were met: a stable operational software state was reached, control of the acoustic seafloor system was confirmed, and the mechanical preparations were finalized.} The latter included the line's unfurling arrangement, the subsea float release mechanism, the sJB rigging, and the preparation of the sub- and mainframe and the spreader cage. \isolate{For the acoustic system, in-air testing confirmed not only direct ranging between transceiver and pinger units, but also the relaying of information and ranging between two autonomous units, which will ultimately allow the geometry of the acoustic seafloor units to be constrained in situ.}

Finally, a subsea preparation and layout plan is presented that highlights the planned P-ONE infrastructure and its installation to the Cascadia Basin in the Pacific Ocean. Through a dedicated extension assembly, the P-ONE infrastructure will connect to the Cascadia node of the NEPTUNE cable network operated by ONC. Dedicated deployment plans for the extension assembly, P-ONE-1, and the acoustic system are described.

Several optimizations and improvements were identified for future lines during the P-ONE-1 integration and testing cycle. These include better control of the ambient humidity during handling of 3D-printed components to reduce the moisture content in the modules; rigorous mechanical testing schedules to identify structural weaknesses early; mechanical fixtures that ease cable management for noise mitigation; and further extended testing of the fully integrated system to prepare the software required for deployment and operation. These lessons directly inform the design of the lines for the next phase of the experiment, already in preparation: the P-ONE Demonstrator.

The work described in this article demonstrates intense and successful planning, integration, and testing cycles of the integrated P-ONE-1 system, and it identifies improvements for future P-ONE lines. P-ONE-1 is now ready for deployment, awaiting mobilization for sea operations. Once deployed, connected, and unfurled, the line will be commissioned subsea to validate all data-taking modes and establish a stable operational state, with a focus on optical background rates, acoustic calibration, and measurements of water and detector properties.

\section*{AI usage statement}
Generative AI tools (large language models) were used during the preparation of this manuscript but only to assist with language editing and paragraph-level copyediting, including improving clarity, concision, and grammar. All technical content, analysis, and conclusions are the sole work of the authors, who reviewed and take full responsibility for the final text.

\section*{Acknowledgments}
We thank Ocean Networks Canada for the very successful operation of the NEPTUNE observatory, as well as the support staff from our institutions without whom this experiment and P-ONE could not be operated efficiently. We acknowledge support from the Canadian Department of Fisheries and Oceans. We acknowledge the support of the Natural Sciences and Engineering Research Council of Canada (NSERC) and the Canadian Foundation for Innovation (CFI). This research was enabled in part by support provided by the BC and Prairies DRI and the Digital Research Alliance of Canada (alliancecan.ca). This research was undertaken thanks in part to funding from the Canada First Research Excellence Fund through the Arthur B. McDonald Canadian Astroparticle Physics Research Institute. P-ONE is supported by the Collaborative Research Centre 1258 (SFB1258) and the Cluster of Excellence ORIGINS, funded by the Deutsche Forschungsgemeinschaft (DFG) under Germany's Excellence Strategy, and by the European Research Council (ERC) under the European Union's Horizon Europe programme through an ERC Advanced Grant. We acknowledge support by the National Science Foundation. This work was supported by the Science and Technology Facilities Council, part of the UK Research and Innovation, and by the UCL Cosmoparticle Initiative. This work was supported by the Polish National Science Centre (NCN).\\

\noindent The authors also thank the IceCube Collaboration for providing access to the testing setup used for the initial PMT characterization in Aachen, and for support from C.~Wendt and A.~Karle regarding the PMT high-voltage base design and PMT choices. We are further grateful to Paul Stoodley and the team at Granville Island Boatyard for their support in preparing and conducting the P-ONE-1 submersion test.

% BIB
\printbibliography
\end{document}